\documentclass[onecolumn]{aa}
\usepackage{graphicx}
\usepackage{txfonts}
\usepackage{natbib}
\usepackage{array}
\usepackage{lineno}
\begin{document}

\title{An adaptive-binning method for generating constant-uncertainty/constant-significance light curves with {\sl Fermi}-LAT data}
\author{B.~Lott\inst{1,2} \and L.~Escande\inst{1,2} \and S.~Larsson\inst{3,4,5} \and J.~Ballet\inst{6}}
\institute{Univ. Bordeaux, CENBG, UMR 5797, F-33170 Gradignan, France
\and
CNRS, IN2P3, CENBG, UMR 5797, F-33170 Gradignan, France
\and 
Department of Physics, Stockholm University, AlbaNova, SE-106 91 Stockholm, Sweden \and The Oskar Klein Centre for Cosmoparticle Physics, AlbaNova, SE-106 91 Stockholm, Sweden \and Department of Astronomy, Stockholm University, SE-106 91 Stockholm, Sweden \and Laboratoire AIM, CEA-IRFU/CNRS/Universit\'e Paris Diderot, Service d'Astrophysique, CEA Saclay, 91191 Gif sur Yvette, France}

\abstract{}{We present a method  enabling the creation of constant-uncertainty/constant-significance light curves with the data of the {\sl Fermi}-Large Area Telescope (LAT).  The adaptive-binning method  enables more information to be encapsulated within the light curve than with the fixed-binning method. Although primarily developed for blazar studies, it can be applied to any sources.}{This method allows the starting and ending times of each interval to be calculated in a simple and quick way during a first step. The reported mean flux and spectral index (assuming the spectrum is a power-law distribution) in the interval  are calculated via the standard LAT analysis during a second step. }{The absence of major caveats associated with this method has been established by means of  Monte-Carlo simulations. We present the performance of this method in determining  duty cycles as well as power-density spectra relative to the traditional fixed-binning method.}{}


\titlerunning{Adaptive-binning method for Fermi-LAT light curves}
\maketitle

\section{Introduction}

Variability studies of blazars provide a wealth of information on the dynamical  processes at work in these objects and yield important 
constraints on their physical parameters. Thanks to its all-sky coverage in 3 hours and large sensitivity, the {\sl Fermi}-Large Area Telescope (LAT),  is enabling a long-term view of variability in the
energy band 0.1--300 GeV for a large sample of $\gamma$-ray blazars. About one thousand blazars are included in the Second LAT Active Galaxy Nuclei Catalog \citep{2LAC}.   
Due to the large variations in flux shown by blazars over different timescales,
producing light curves that preserve the information provided by the data is not an easy task. With a regular (fixed) time binning, using long bins will smooth out the fast 
variations assessable during bright flares. Conversely, using short bins might lead to hard-to-handle upper limits during low-activity periods. Here we present a 
simple method where the bin width is adjusted by requiring a constant relative flux uncertainty, $\sigma_{\ln\,F}=\Delta_0$, (alternatively a
constant significance) in each bin. This method allows for more 
information to be encapsulated within the light curve than can possibly be achieved for a fixed-binning light curve, without favoring any {\sl a priori} arbitrary timescale and while avoiding upper limits. These advantages come at the expense of an
iterative procedure required to find the proper bin widths satisfying the chosen criterion.  The elaborate, fairly computing-intensive  maximum-likelihood procedure necessary to analyze LAT
data does not lend itself very well to such an iterative procedure. While light curves with constant signal-to-noise ratios are commonly used at other wavelengths such as the X-rays, the above difficulty has so far hampered the generation of corresponding light curves in the LAT domain. The purpose of the method presented in this paper (referred to as the adaptive-binning method) is to provide a solution to this problem.

The adaptive-binning method bears some resemblance to the Bayesian-Blocks (BB) method \citep{scargle}. Both methods
propose a time binning which adapts itself to the data instead of an {\sl a priori} arbitrary regular binning. The main difference is that the BB
method gives the most probable segmentation of the observation period into time intervals during which the photon arrival rate is perceptibly constant, i.e., has no statistically significant variations. A bin then defines the period over which a constant flux is observed. In the adaptive-binning method, the segmentation criterion is based on a given value of $\sigma_{\ln\,F}$ (or significance) regardless of whether the flux varies within the bin.
The adaptive-binning method is expected to be more sensitive (i.e. allows the study of fainter sources) than the BB method since it makes use of both the photon spatial and energy information and properly accounts for the diffuse backgrounds while the BB method does not. The flare rise and decay times can also be determined more accurately with the adaptive-binning method. 

This paper presents the method and explores its possible inherent biases/caveats. For the sake of clarity,  only light curves obtained with a constant relative flux uncertainty will be discussed, but changing the criterion to a constant significance, defined from the Test Statistic\footnote{The Test Statistic is defined as $TS$ = 2($\ln \mathcal{L}$(source)$ - \ln \mathcal{L}$(nosource)), where  $\mathcal{L}$ represents the likelihood of the data given the model with or without a source present at a given position in the sky. See Appendix for details.} ($TS$), is straightforward. The method is described in Section 2. Validations with simulations are presented in Section 3. 
Section 4 presents results on timing analyses using the adaptive-binning light curves regarding the determination of the duty cycle and the power density spectra of the $\gamma$-ray source. A summary is given in Section 5.

\section{Description of the method}

\subsection{Optimization of the time bins}
The goal of the method is to compute light curves for an integral flux above a given energy $E_{min}$, with constant $\Delta_0$  in all time bins.
The source energy spectrum  is assumed to be a power-law distribution with photon spectral index $\Gamma$.  Although the method allows any $E_{min}$ to be used,  fluxes are reported in the following above the optimum energy ($E_1$) for which the accumulation times needed to fulfill the above condition  (i.e., bin widths) are the shortest relative to other choices of $E_{min}$ (see the Appendix for the derivation of $E_1$). The energy $E_1$ depends on the signal/background ratio but is independent of exposure. It is calculated here using the flux and $\Gamma$ determined over the whole time range spanned by the data.  

The method consists in solving the following equation for $T_1$ (ending time of the interval) given a starting time $T_0$:
\begin{eqnarray}
\sigma_{\ln\,F}(T_0,T_1,\bar{F},\bar{\Gamma})=\Delta_0 \label{eq:ff} \end{eqnarray} 
\noindent where $\bar{F}$ and $\bar{\Gamma}$ are the average flux and photon spectral index over the interval $[T_0,T_1]$ respectively.
In order to avoid the penalty of excessive computing time required by the standard maximum-likelihood analysis, $\sigma_{\ln\,F}$ is estimated from the time-ordered list of photons contained in the region of interest (ROI), as described in the Appendix (see Eq. \ref{eq:dAi}).  

Since the source is potentially variable, $\bar{F}$ must be evaluated consistently in each time bin. $\bar{F}$ is estimated through a procedure maximizing the likelihood (see Eq. \ref{eq:dlogl}), this procedure being similar to a simplified version of the standard LAT analysis. In principle $\bar{\Gamma}$ should be evaluated similarly, but in practice $\bar{\Gamma}$ can be left fixed  
\footnote{$\Gamma$ can be taken from a source catalog when available (for real data) or set to the measured value over a long period.} to speed up computation. Taking a constant $\Gamma$ is largely justified since spectral variations with time  
have been found to be very moderate in LAT-detected blazars\footnote{Preliminary light curves of bright LAT-detected blazars computed with real data proved that taking a constant photon spectral index in the bin width estimate provides reasonably accurate results in terms of $\sigma_{\ln\,F}$.}  \citep[see ][]{spec}.

In practice, due to the discrete nature of the data, Eq. \ref{eq:ff} can only be solved in an approximate way. The procedure is the following:
\begin{itemize}
\item For a given $T_0$, $\bar{F}$ (initially estimated over the whole period for the first bin) and $\bar{\Gamma}$, the function $\sigma_{\ln\,F}(T_0,T_1,\bar{F},\bar{\Gamma})$ is monotonically decreasing with increasing $T_1$ (see Eq. \ref{eq:dAi}). The detection time of the {\sl earliest} photon leading to the fulfillment of the condition $\sigma_{\ln\,F}<\Delta_0$ is hence taken as $T_1$.
\item $\bar{F}$ is then reestimated over $[T_0,T_1]$ and $\sigma_{\ln\,F}$ is reevaluated using this new $\bar{F}$ value. Then

$\bullet$ if $\sigma_{\ln\,F}$ is equal to  $\Delta_0$ within a predefined tolerance, convergence is achieved and $T_0$ is replaced by $T_1$;

$\bullet$ otherwise, $T_1$ is reevaluated with the updated value of $\bar{F}$. A bisection method, making use of the set of $T_1$ estimates obtained in previous iterations, complements the procedure to speed up convergence. 
\item The whole procedure is repeated until convergence is achieved.
\end{itemize}

The times T$_1$ for the different intervals are calculated sequentially until  the condition $\sigma_{\ln\,F}<\Delta_0$ cannot be fulfilled using the remaining set of
photons (these photons are left unused). Once the whole set of time intervals has been determined (this stage is referred to as ``Step 1'' in the following), $\bar{F}$, $\bar{\Gamma}$ and $\sigma_{\ln\,F}$ are recalculated for each interval with the standard {\sl pylikelihood} analysis (``Step 2''). The consistency between these values of $\sigma_{\ln\,F}$ and $\Delta_0$ is then checked. Using Monte-Carlo simulations, an excellent agreement has been observed in most cases, making further iterations unnecessary. 

Being driven by the number of accumulated source photons, the bin widths provided by this method depend on both the source flux and the exposure of the instrument for the source location. The modulation of the LAT daily exposure over the precession period (53.4 days) ranges between a few percent and $\pm50$\% (for a rocking angle of 50$^\circ$) depending on the source declination.  

\subsection{{\sl Pylikelihood} analysis}

Results described in the following were obtained via an unbinned {\sl pylikelihood} analysis performed with the {\sl Fermi}-LAT ScienceTools software package\footnote{http://fermi.gsfc.nasa.gov/ssc/data/analysis/documentation/Cicerone/} version v9r19p0.
The tools used in Step 1 and the {\sl pylikelihood} analysis made use of the same list of photons with energy above 100 MeV and contained within a 20$^\circ$-in-diameter ROI.  The P6\_V3\_DIFFUSE set of instrument response functions (IRFs) was employed.   The Galactic 
diffuse emission model version "gll\_iem\_v02.fit"  and the isotropic (sum of residual instrumental and extragalactic diffuse) background ``isotropic\_iem\_v02.txt'' 
(files provided with ScienceTools) were used throughout the procedure. The normalization of both diffuse components were left free in the fit for the {\sl pylikelihood} analysis. For uniformity over many realizations produced under the same conditions, the optimum energy was evaluated from the true flux and spectral index used in Monte-Carlo simulations.

\section{Validation with simulations}

In this section, we first present the results of the adaptive-binning method  applied to simulated variable sources. Then, we address some of the possible caveats of this method  by considering 
simulated constant sources. For this purpose, we compare results obtained for adaptive-binning and fixed-binning light curves regarding the following issues: biases and fluctuations in the measured fluxes and photon indices, correlation between flux and $\Gamma$ and finally the flux correlation between consecutive bins. Fixed-binning analyses were carried out with the same total number of bins as the adaptive-binning analyses to enable meaningful comparisons. The criterion adopted for all adaptive-binning analyses is $\Delta_0$=25\%.

\subsection{Description of simulations}
Data were simulated  in the energy range 0.1--200 GeV with the tool {\sl gtobssim}, part of the {\sl Fermi}-LAT ScienceTools software package,
which generates photon events from astrophysical sources and models their detections according to the IRFs. The P6\_V3\_DIFFUSE set of IRFs  was used consistently with the analysis described above. These simulations do not include all known systematic effects of the real data, in particular related to the varying instrumental background during the orbit. However, this lack is not believed to significantly affect the conclusions drawn thereafter due to the moderate magnitude of these effects  \citep[a few \% in the flux, ][]{Crab}. The simulations were performed for an 11-month long period starting at the beginning of the scientific operation of the {\sl Fermi} mission, using the actual LAT pointing history. Although light curves shown in the following use fluxes above the optimum energy, the different examples are distinguished according to their true fluxes above 100 MeV (F$_{100}$ in $10^{-7}$ ph cm$^{-2}$ s$^{-1}$) to make comparisons clearer. All sources  were simulated at the same high-Galactic latitude sky position, (l,b)$\simeq$(80$^\circ$, 80$^\circ$).

\subsection{Light curves of variable sources}

In this section, the method is illustrated for three variable sources which were simulated by means of templates showing realistic flux variations typical of blazars. The source spectra were modeled with a single power-law function with an average flux $F_{100}$=2 (corresponding to a fairly bright blazar detected by the LAT) and a fixed photon index $\Gamma$=2.4. The optimum energy is $E_1$=230 MeV for these parameters.  

\begin{figure}[ht!]

 \centering

 \includegraphics[scale=0.55]{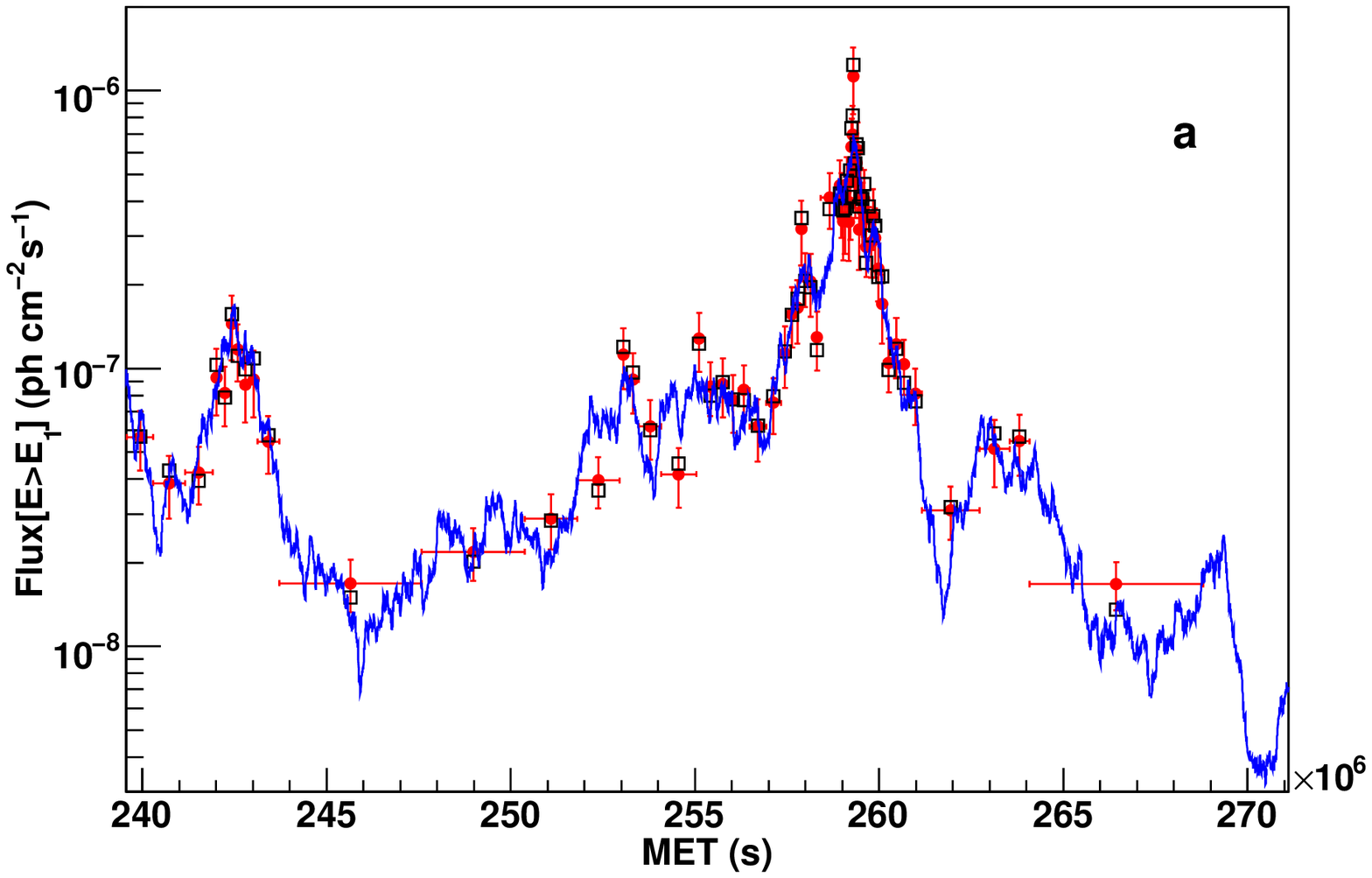}
 \includegraphics[scale=0.55]{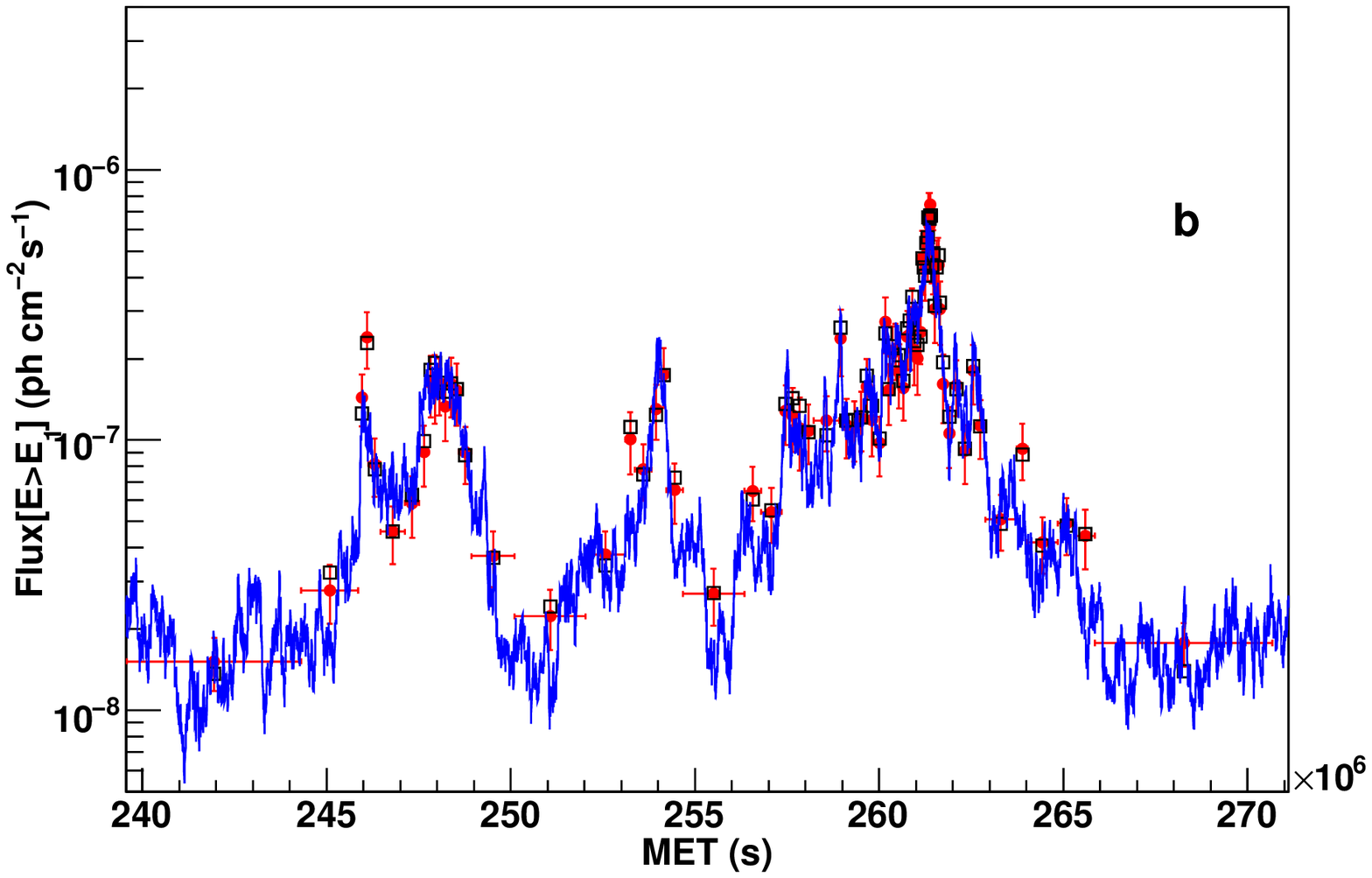}
 \includegraphics[scale=0.55]{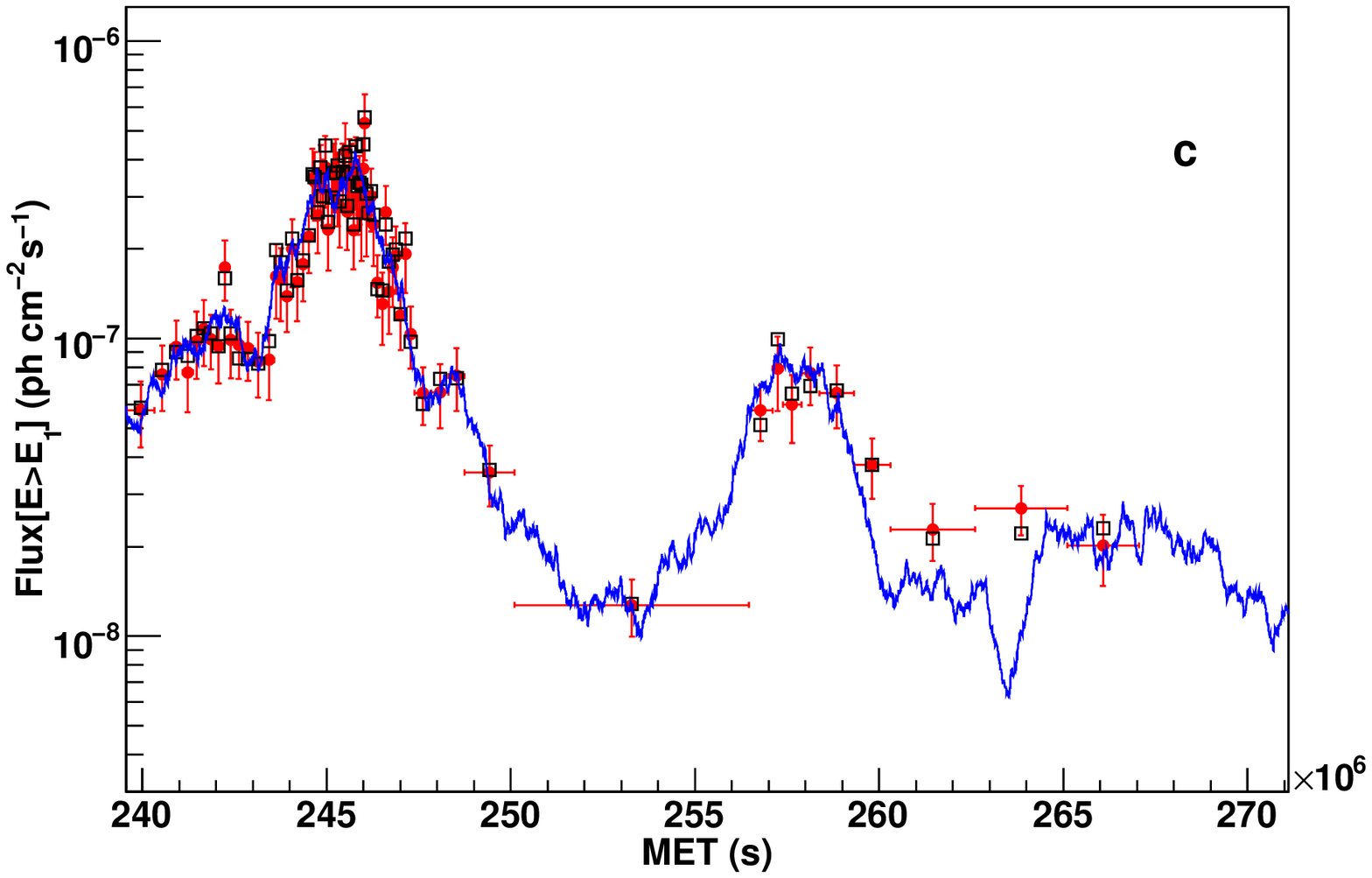}
  \caption{Simulated light curves (blue) with the adaptive-binning light curves superimposed (Step 1: black open squares, Step 2: red solid circles) for three variable sources. MET stands for Mission Elapsed Time.}

  \label{fig:lc_var}

\end{figure}

\begin{figure}[ht!]

 \centering

 \includegraphics[scale=0.55]{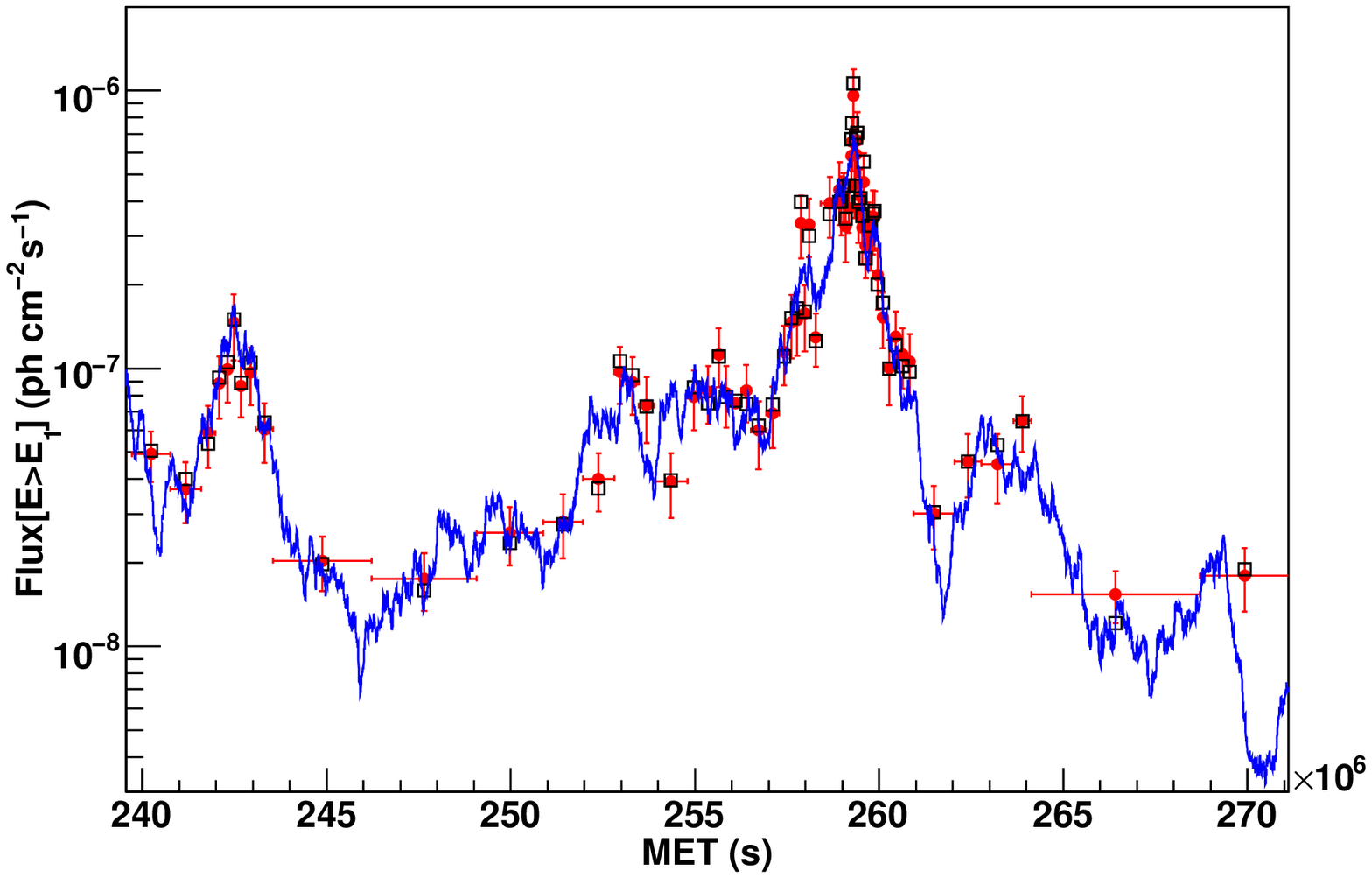}
  \caption{Same as Fig. \ref{fig:lc_var}a but the analysis was conducted assuming a reversed arrow of time.}

  \label{fig:lc_var_rev}

\end{figure}

Figure \ref{fig:lc_var} compares the true light curves with light curves calculated using the adaptive-binning method  for the three sources. The analysis for the light curve shown in Figure \ref{fig:lc_var} top was also conducted assuming a reversed arrow of time, the results being presented in Figure \ref{fig:lc_var_rev}. Statistically significant fluxes can be determined over very different time scales, ranging from a few hours to several months in these examples.   The occurrence of significant short-term flaring activity during the low-flux states can be ruled out.

An adverse feature of the method is the flare truncation (the light curve in Figure \ref{fig:lc_var}c shows such a feature at MET$\simeq$255000000). When a significant flare 
occurs after a long period of low activity, a non-negligible number of photons from the flare has to be "consumed" to counterbalance the accumulated background photons and enable the criterion to be fulfilled. Possible ways to mitigate this effect are being explored and will be implemented in future versions.  The tools allow light curves to be generated by considering either increasing or decreasing photon times, so the robustness of any analysis results (time lags, duty cycles, rise and decay times...) can be tested by using both flavors.   

Figure \ref{fig:lc_var}c may convey the impression that the adaptive-binning method will lead to light curves lagging the real one. However,  cross-correlation analyses have shown that no significant lag is observed for the light curves investigated here, probably because truncation affects only very few bins and the flare leading edges are still well sampled.

\begin{figure}[!h]

 \centering

 \includegraphics[scale=0.3]{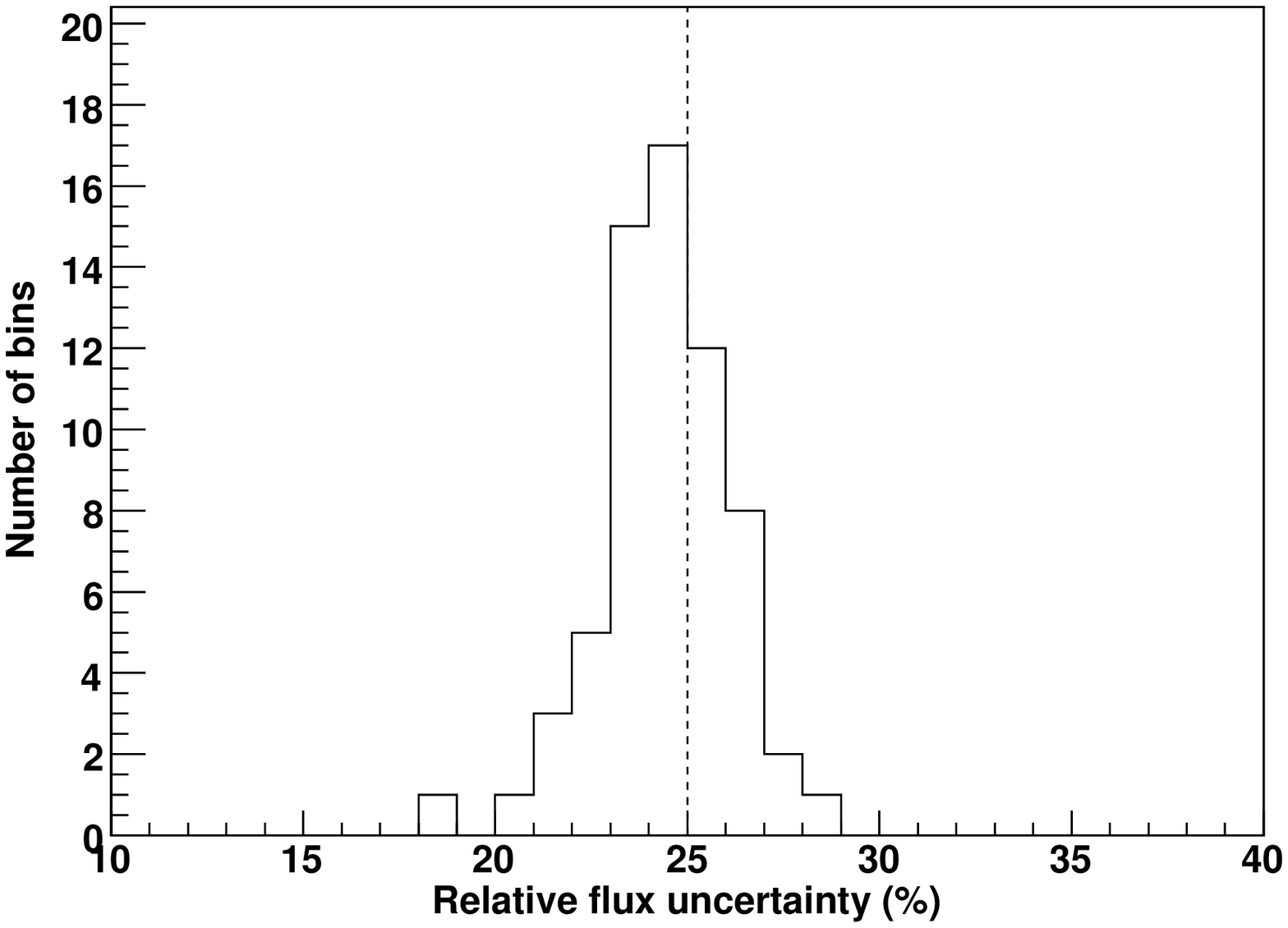}
 \includegraphics[scale=0.3]{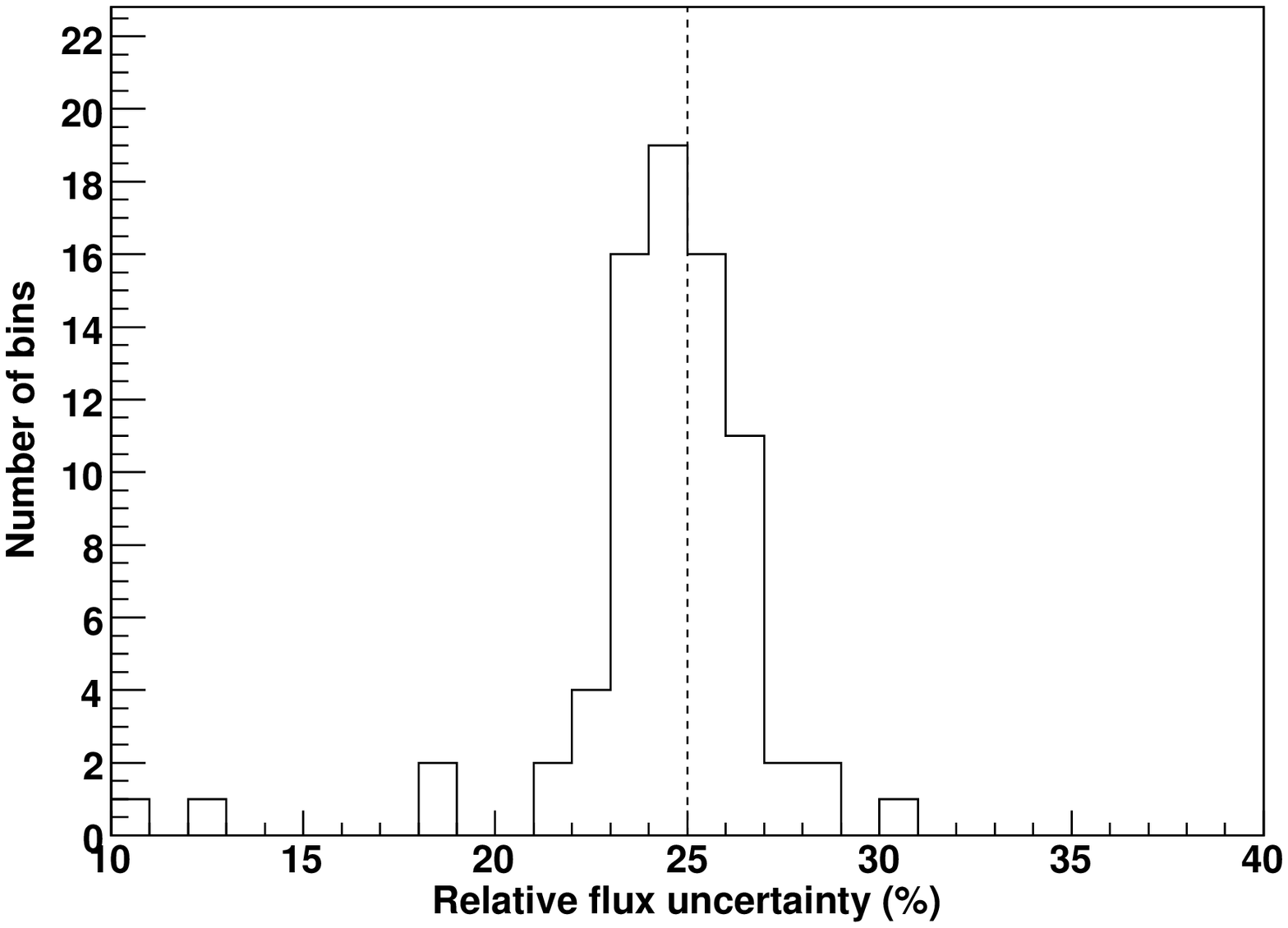}
 \includegraphics[scale=0.3]{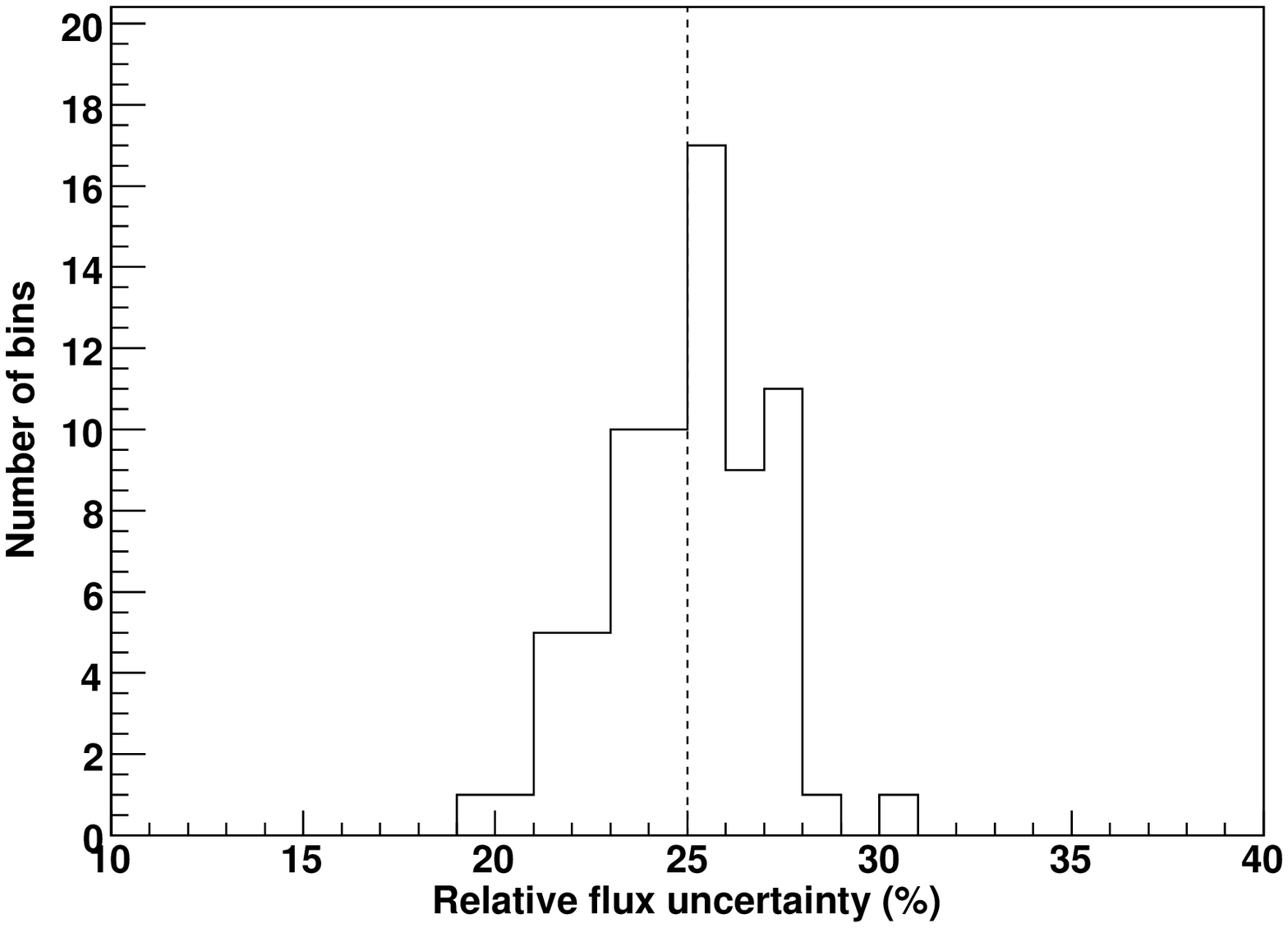}
  \caption{Distribution of relative flux uncertainties given by  the {\sl pylikelihood} analysis for three variable sources. The dashed line depicts the target $\Delta_0$=25\%.}

  \label{fig:perf_var}

\end{figure}

Figure \ref{fig:perf_var} shows the distribution of relative flux uncertainties calculated from the {\sl pylikelihood} analysis for the three variable sources. All three distributions are centered at a value close to the target $\Delta_0$=25\%  (depicted by the dashed line) with essentially all values included in the range 20\%--30\% with a typical rms of  3\%, demonstrating the performance of the method.

\subsection{Caveats explored with steady sources}

\begin{figure}[!h]

 \centering

 \includegraphics[scale=0.65]{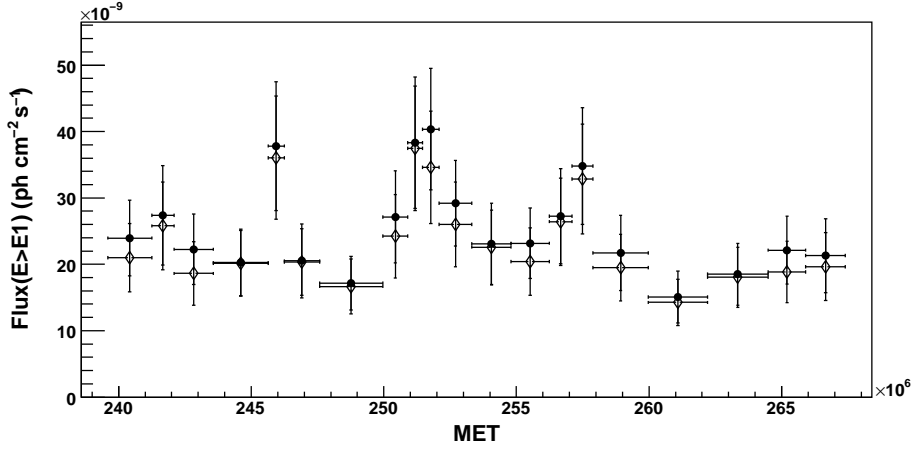}
  \caption{Comparison of flux from the tool, i.e., after Step 1 (open diamonds) and flux resulting from the {\sl pylikelihood} analyses, i.e., after Step2  (filled circles) for a simulated constant source  with $F_{100}$=1 and $\Gamma$=2.4.}

  \label{fig:est_gt}

\end{figure}

\begin{figure}[!h]

 \centering

\includegraphics[scale=0.45]{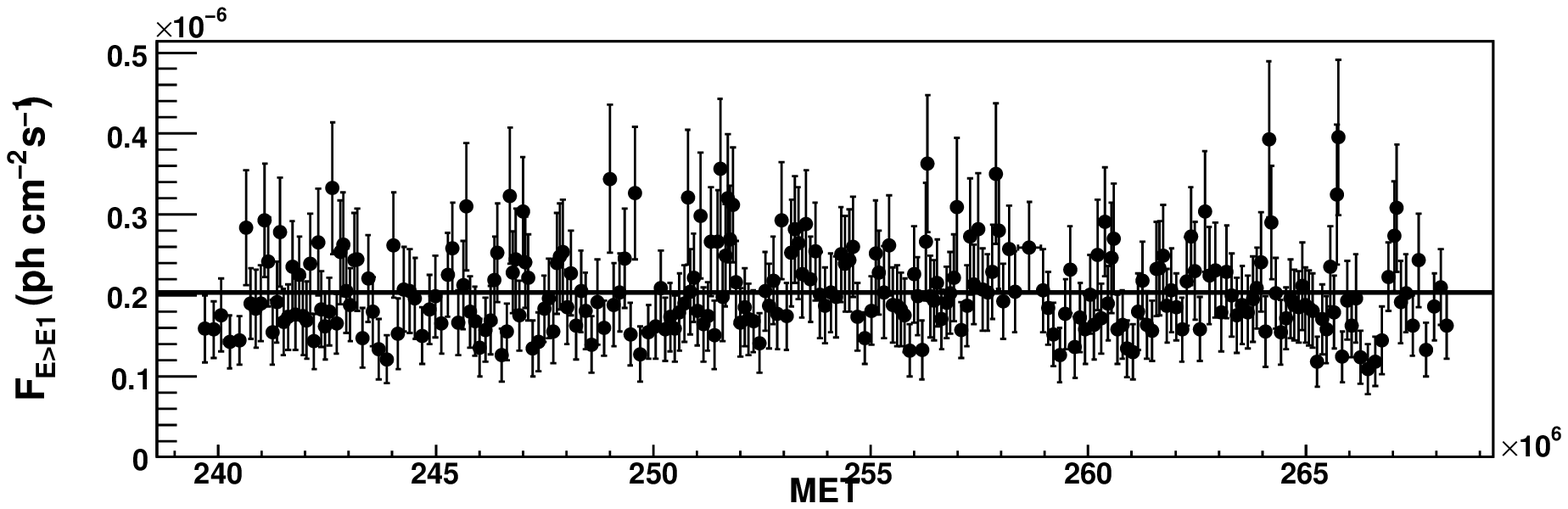}
 \includegraphics[scale=0.45]{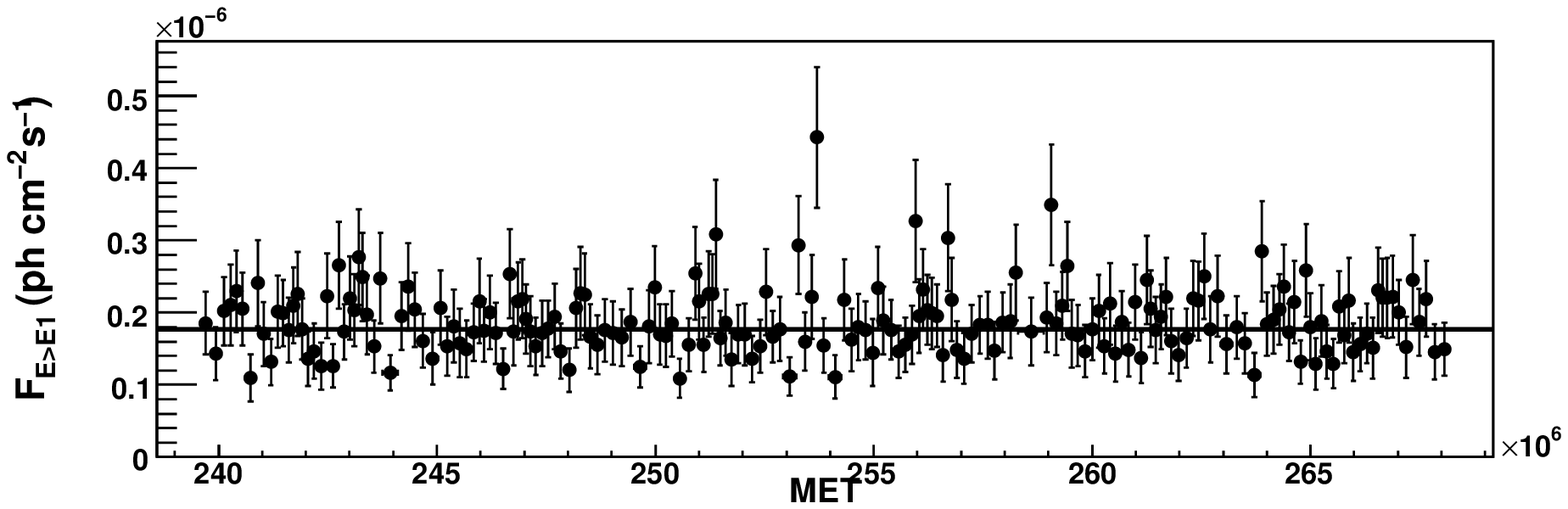}
 \includegraphics[scale=0.45]{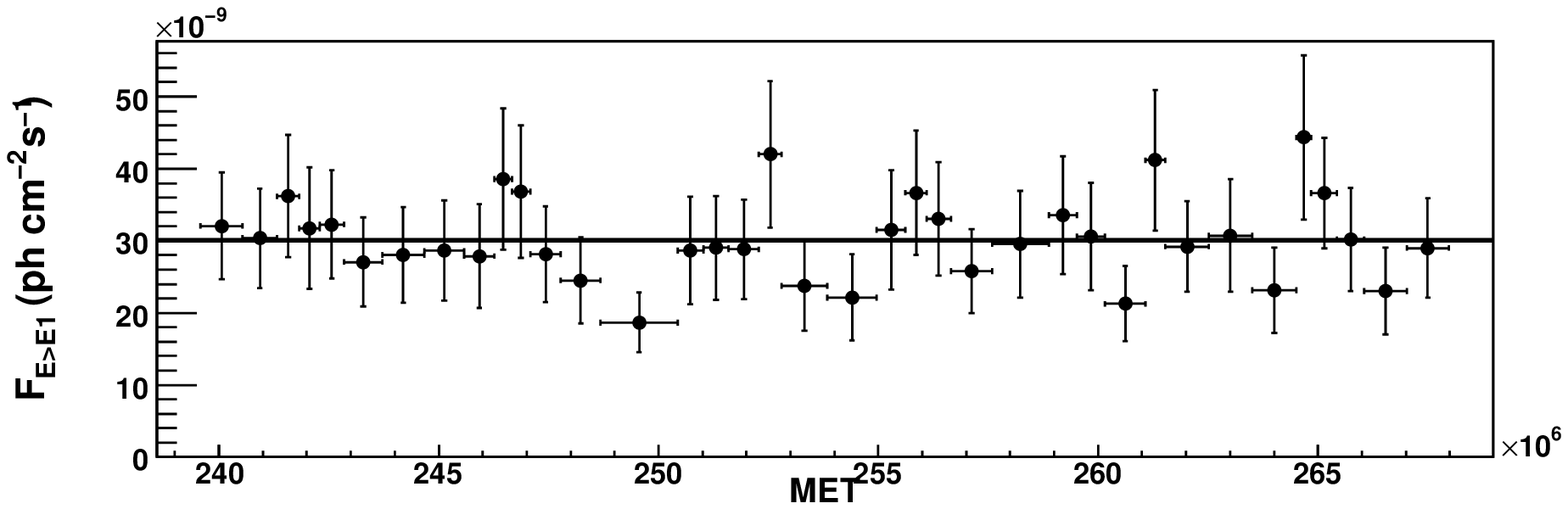}
 \includegraphics[scale=0.45]{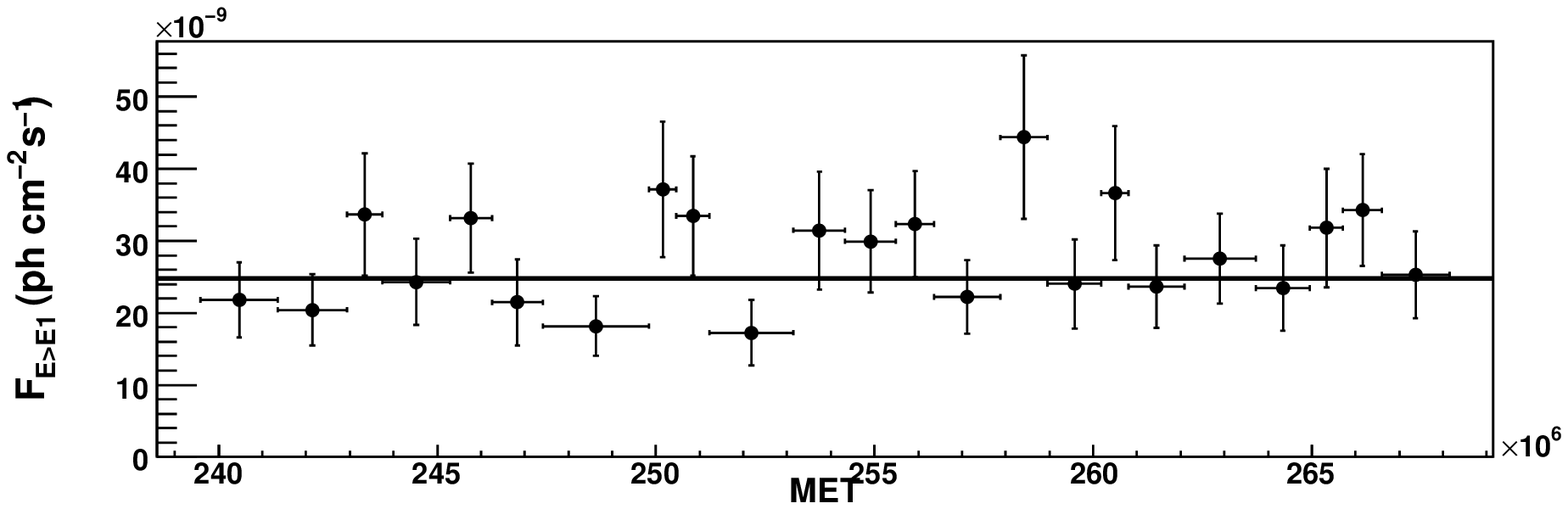}
 \includegraphics[scale=0.45]{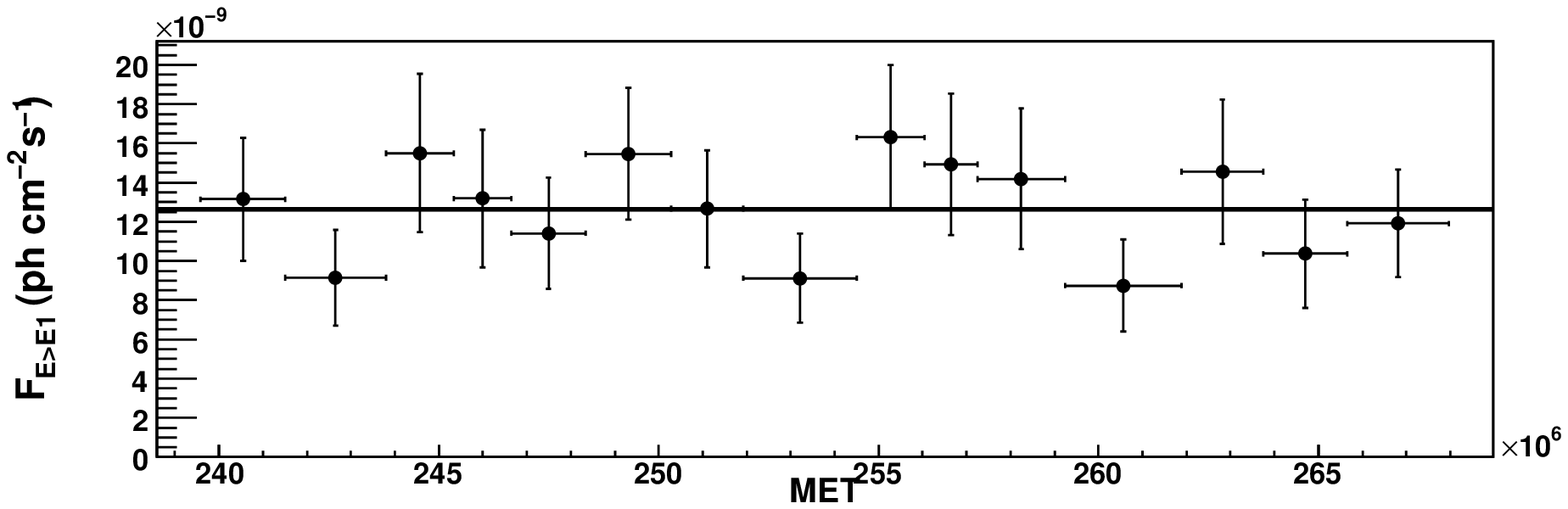}
 \includegraphics[scale=0.45]{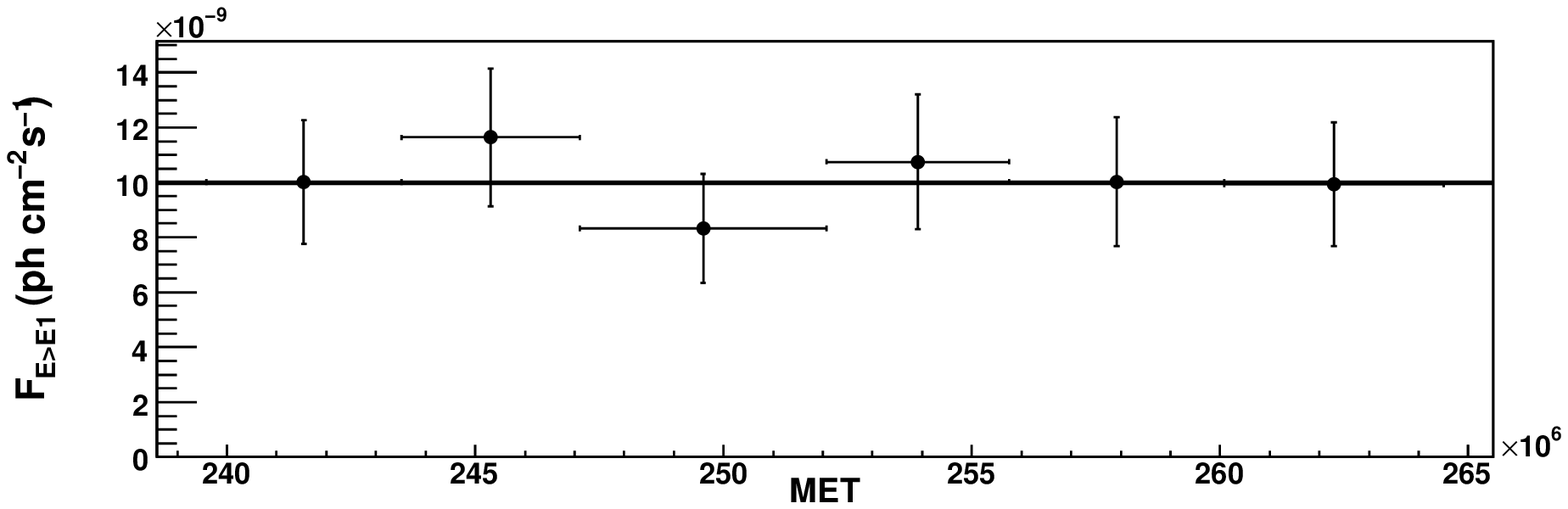}
 
  \caption{Examples of light curves for the different sets of parameters : $\Gamma$=2.0 (left) and $\Gamma$=2.4 (right), $F_{100}$=5, 1, 0.5  from top to bottom. The solid line depicts the true flux value.}

  \label{fig:lc_ex}

\end{figure}

\begin{figure}[ht!]

 \centering

 \includegraphics[scale=0.45]{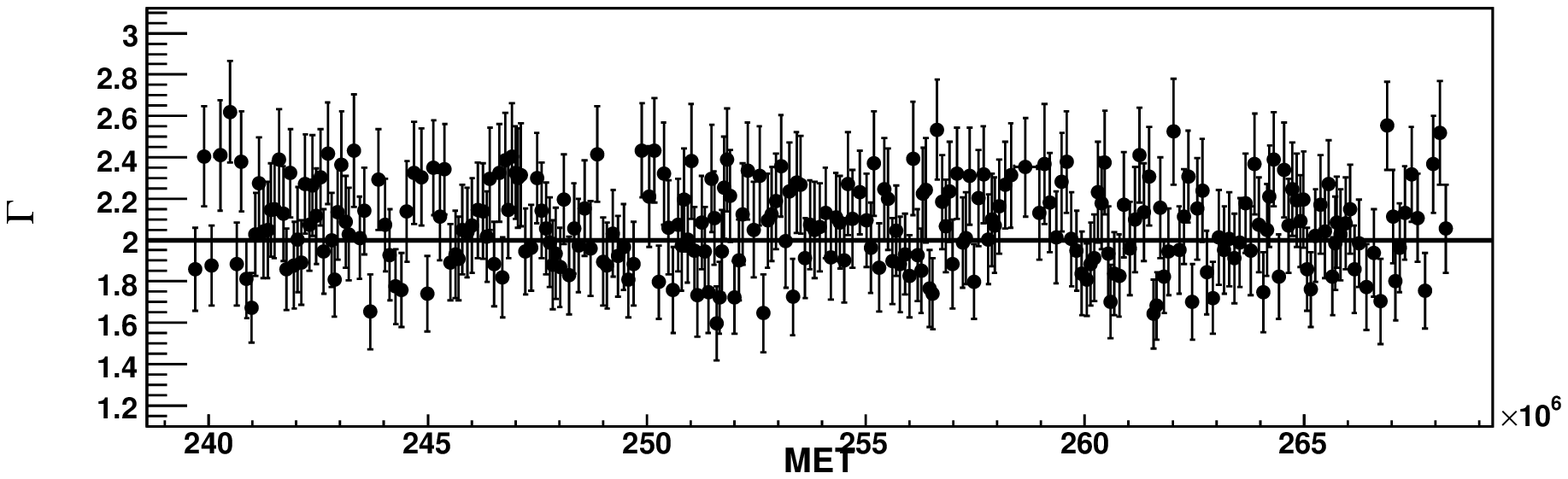}
 \includegraphics[scale=0.45]{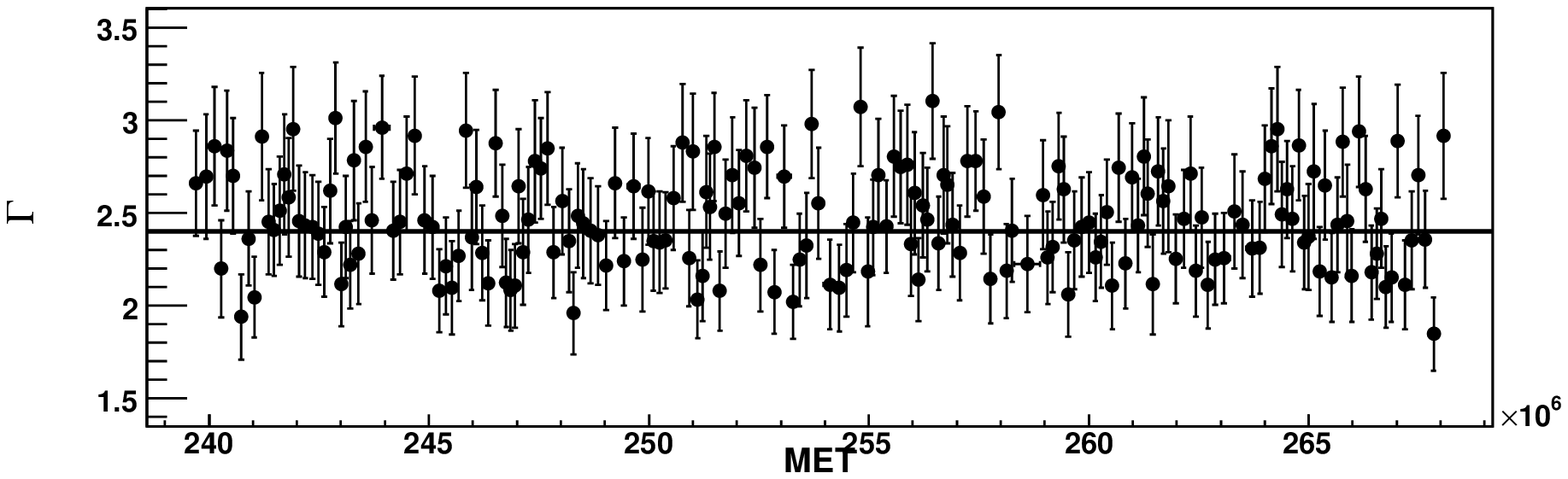}
 \includegraphics[scale=0.45]{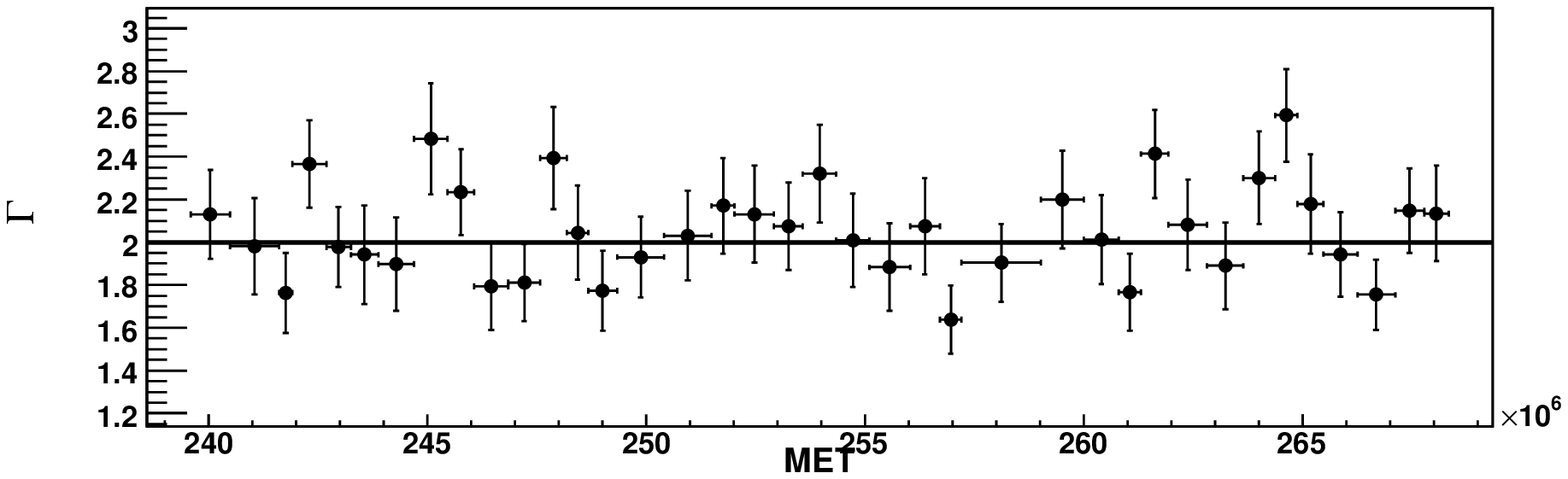}
 \includegraphics[scale=0.45]{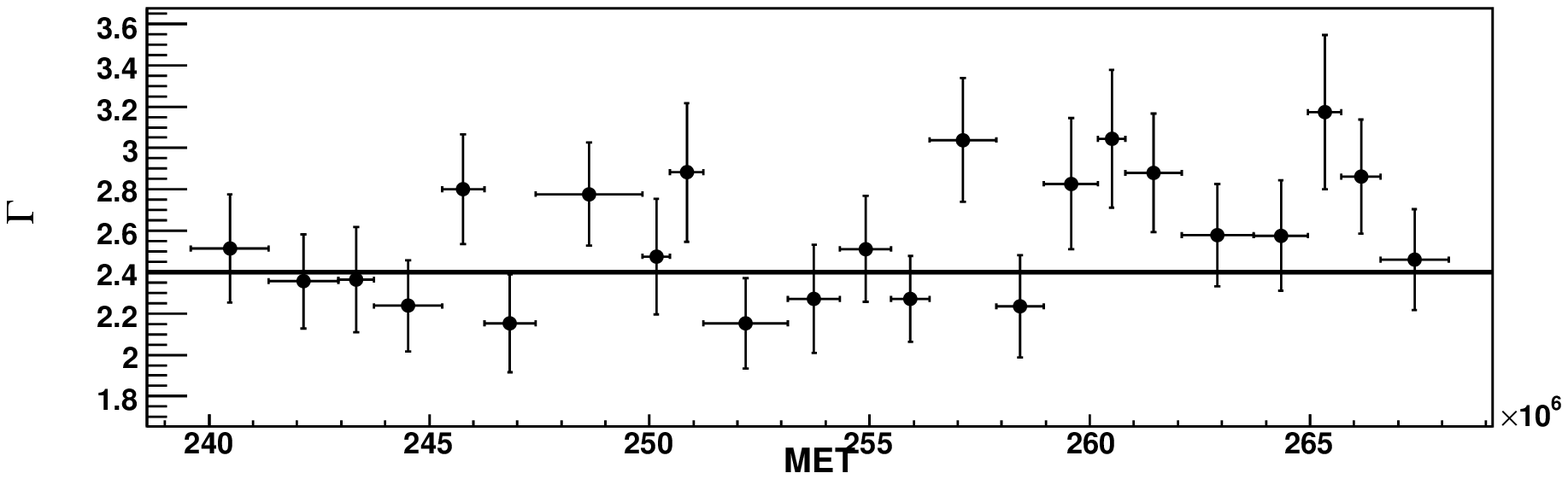}
\includegraphics[scale=0.45]{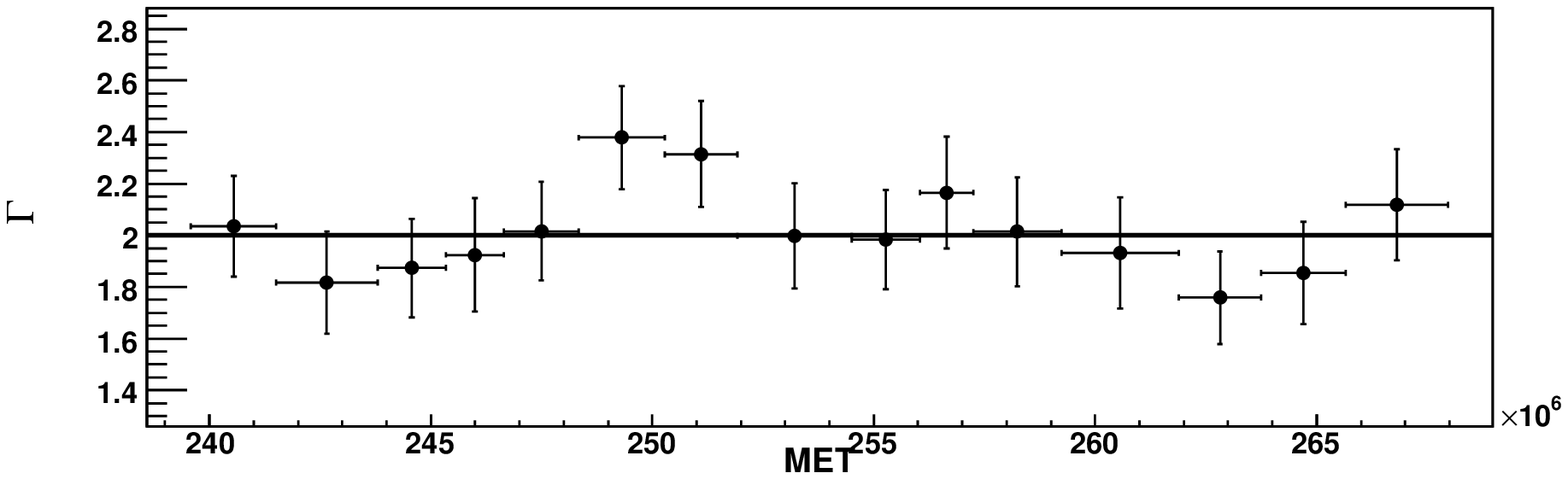}
 \includegraphics[scale=0.45]{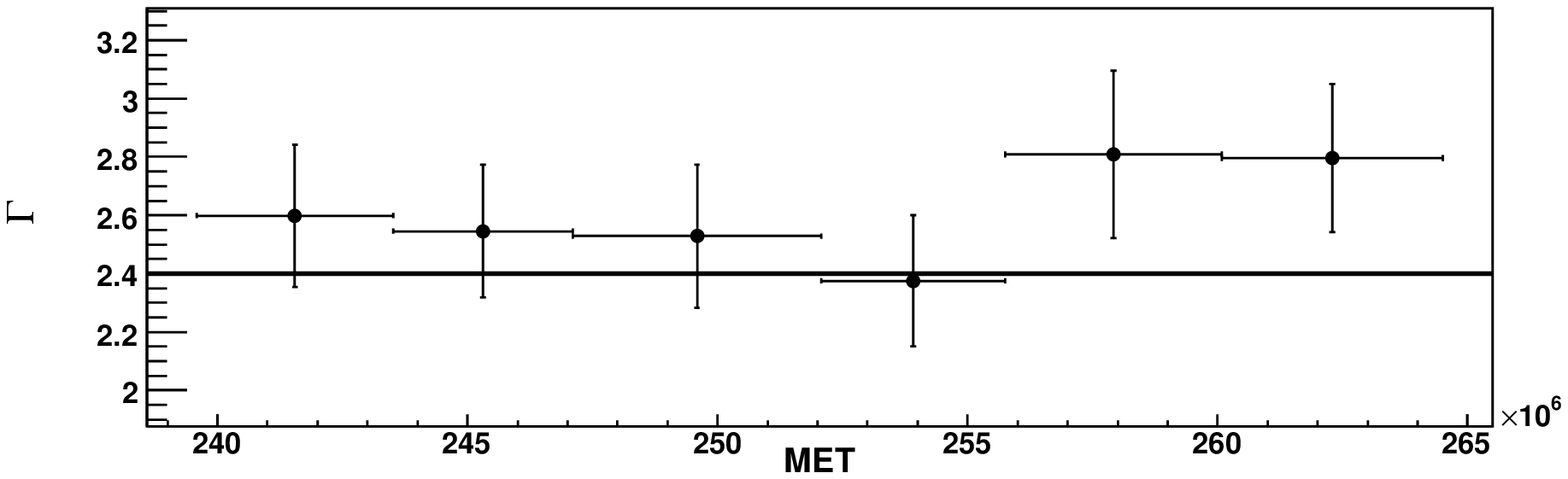}
  \caption{Temporal evolution of the measured photon spectral index: $\Gamma$=2.0 (left) and $\Gamma$=2.4 (right), $F_{100}$=5, 1, 0.5  from top to bottom. The solid line depicts the true photon index.}

  \label{fig:lc_index_ex}

\end{figure}

\begin{figure}[ht!]
\centering

\includegraphics[scale=0.4]{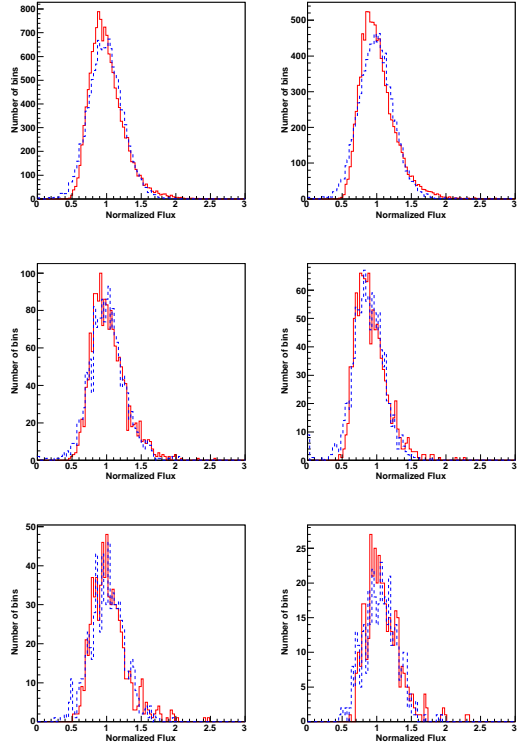}
\caption{Normalized flux distributions for adaptive (solid) and fixed (dashed) binnings and steady sources with $\Gamma$=2.0 (left) and $\Gamma$=2.4 (right), $F_{100}$=5, 1, 0.5  from top to bottom}

\label{fig:lc_flux_dist}

\end{figure}

Steady sources with three different fluxes ($F_{100}$=0.5, 1 and 5) 
and two different photon spectral indices ($\Gamma$= 2.0 and 2.4) were simulated over the same 11-month long period as for variable sources. Fifty realizations were performed for each case.  The average numbers of bins in the light curves as well as the optimum energies are reported in Table \ref{tab:npts} for the different cases. For $F_{100}$=1 and $\Gamma$=2.0, a light curve where fluxes are estimated after Step 1 (i.e., using the simple maximum-likelihood procedure described in the Appendix) for a particular realization  is compared to that resulting from Step 2 (i.e., where fluxes are computed  with the whole {\sl pylikelihood} analyses with the time intervals established in Step 1) in  Figure \ref{fig:est_gt}. There is a good agreement between these fluxes, which holds for all realizations and parameter sets. 
Examples of adaptive binning light curves for the flux are shown in Figure \ref{fig:lc_ex} for the different cases, while 
light curves for the  photon spectral index are displayed in Figure \ref{fig:lc_index_ex}. Figure \ref{fig:lc_flux_dist} compares the flux distributions obtained via the adaptive- and fixed-binning methods by summing up all realizations. For a given set of conditions, the distributions have similar means and rms but different skews. A positive skew (typically equal to 1 $\sigma$ for the distributions shown in Figure \ref{fig:lc_flux_dist}) is a feature of the adaptive-binning method. A positive fluctuation in flux leads to a shorter-than-average bin associated with a high apparent flux. A negative fluctuation has less chance to be recorded since the bin will be extended until enough photons have been accumulated to meet the $\Delta_0$-wise criterion. 
%The cumulative distribution functions (CDF), where the fluxes obtained with the adaptive-binning method were weighted with the bin widths,  are compared in Figure \ref{fig:lc_cdf}. 
The photon index distributions have been found to be very similar for the two methods and close to gaussian distributions. 

\begin{figure}[!h]

 \centering

 \includegraphics[scale=0.15]{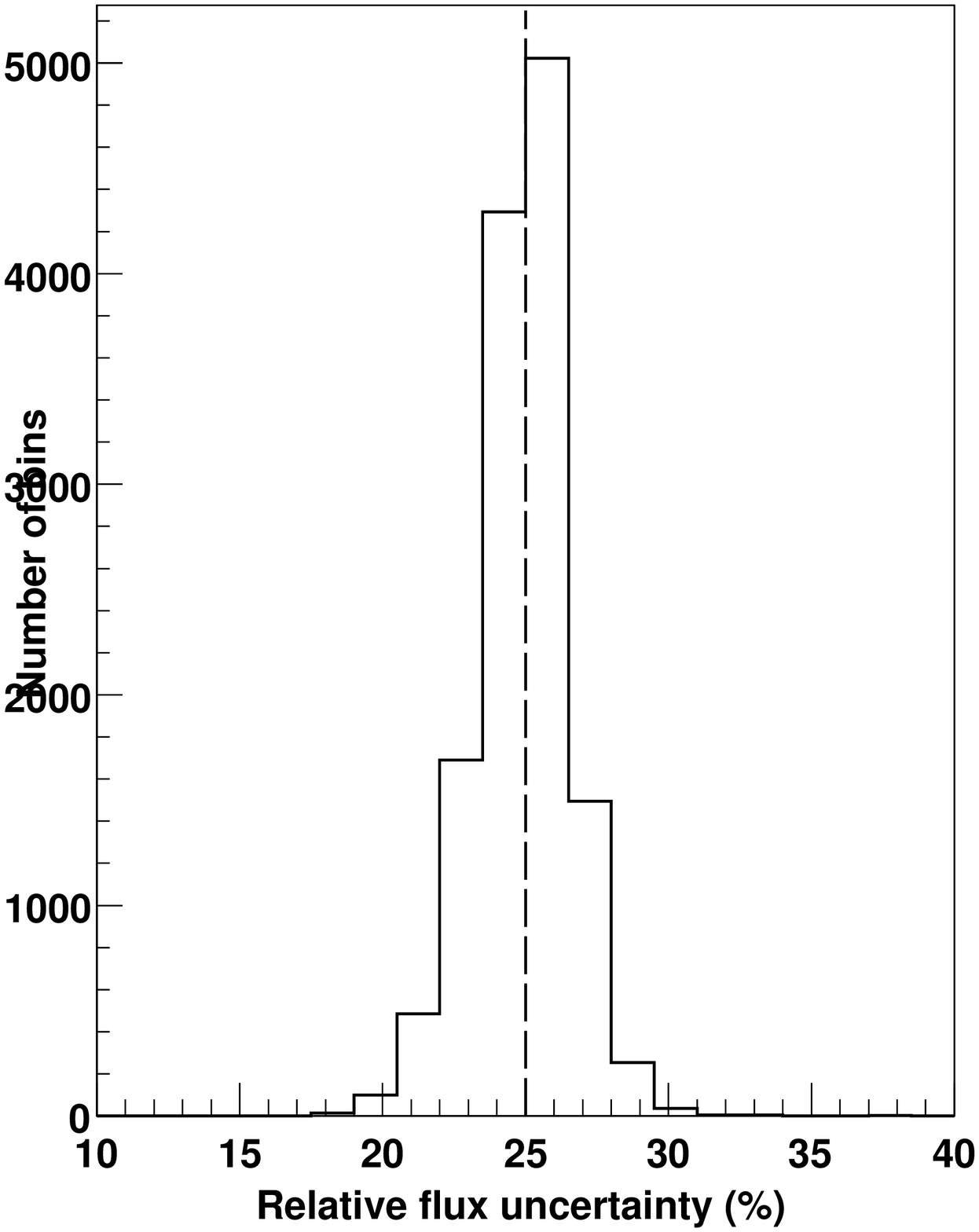}
 \includegraphics[scale=0.15]{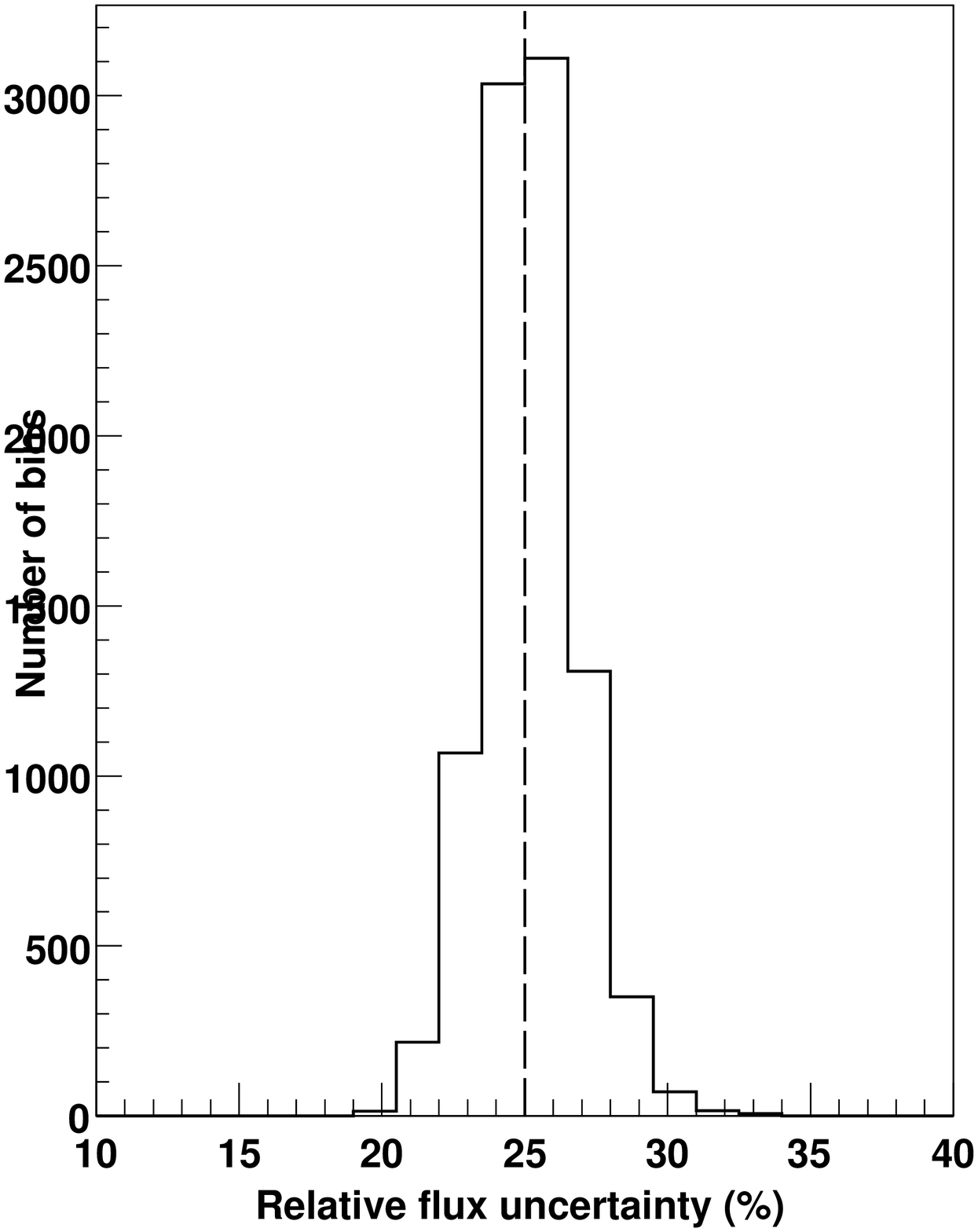}\\
 \includegraphics[scale=0.15]{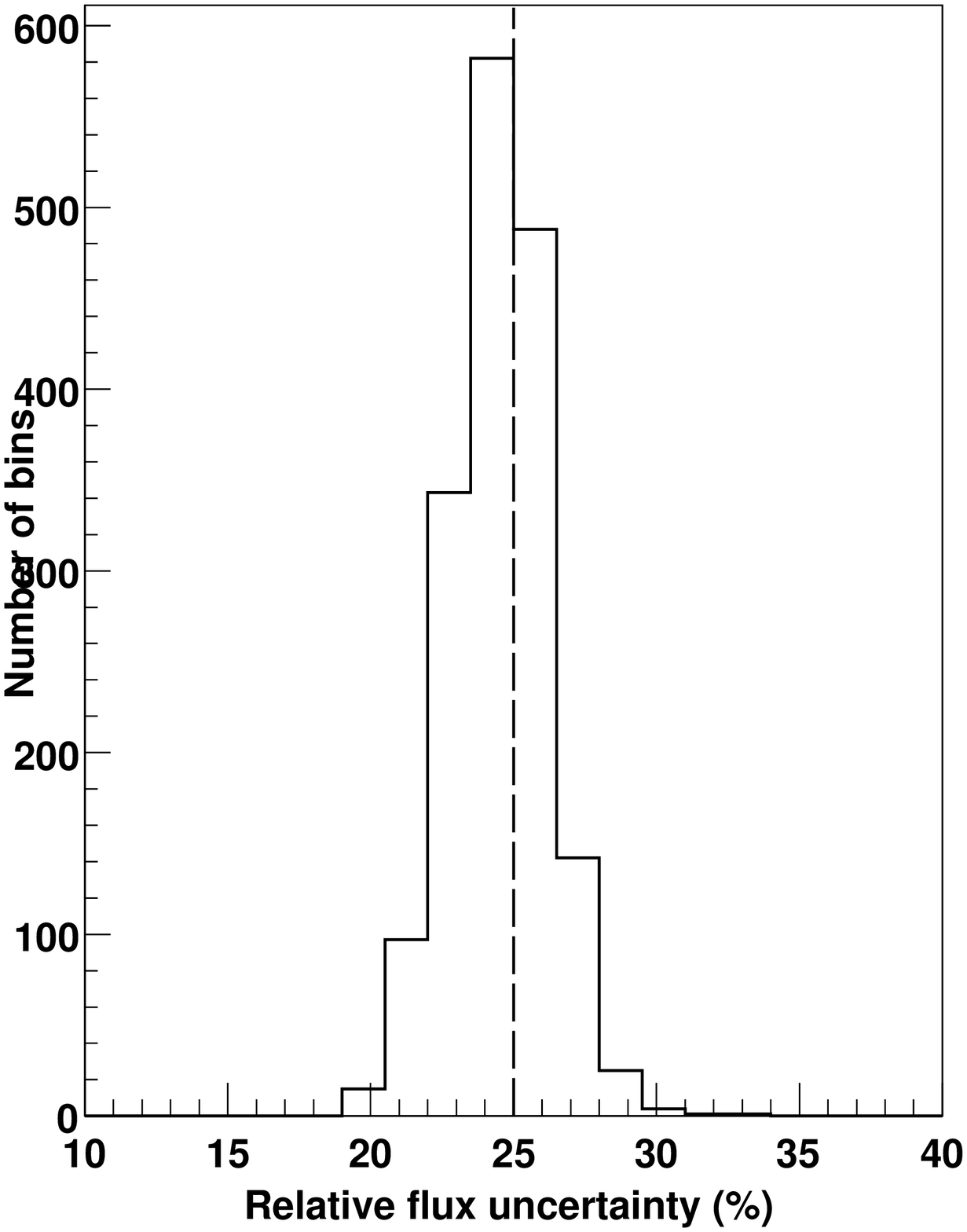}
 \includegraphics[scale=0.15]{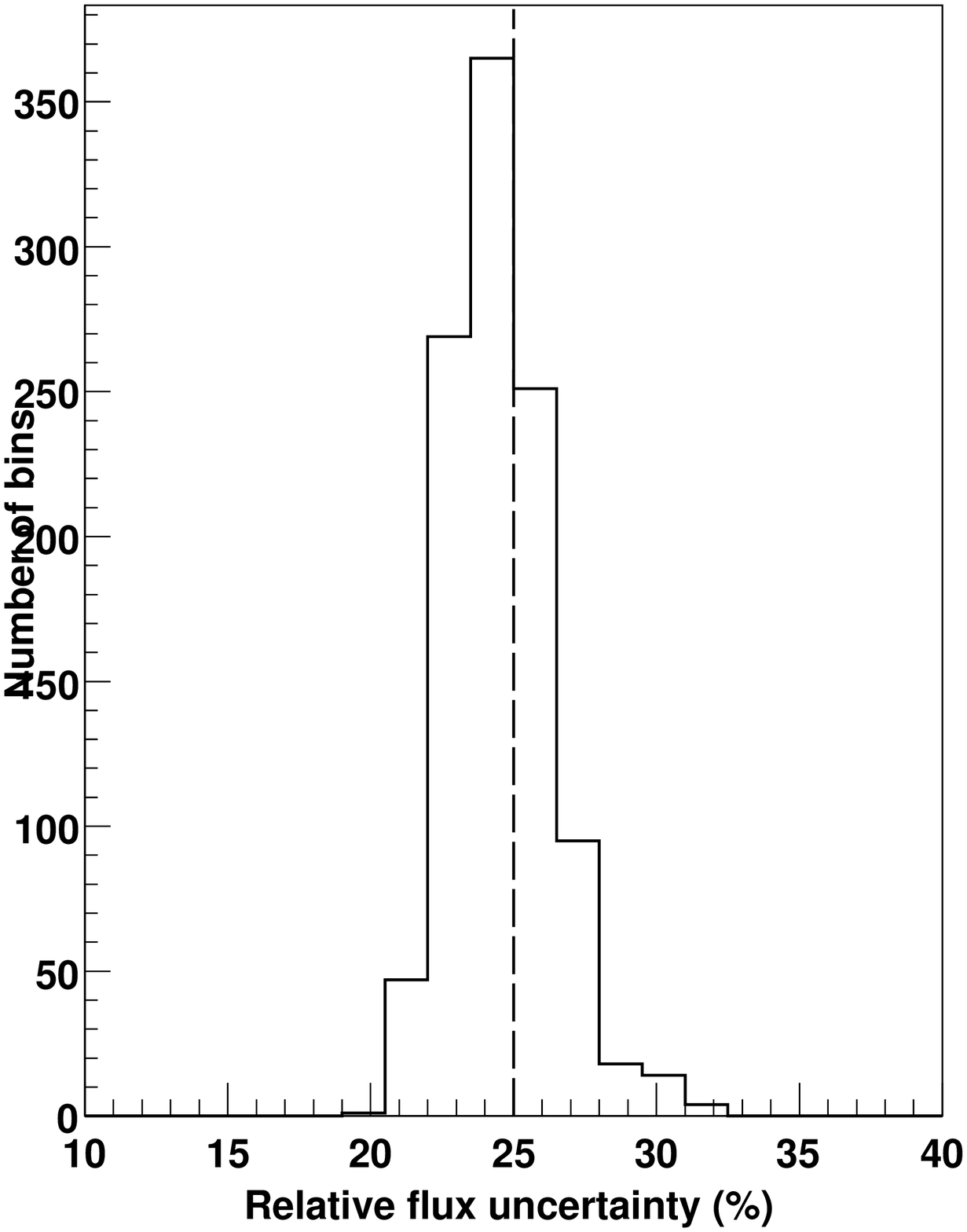}\\
 \includegraphics[scale=0.15]{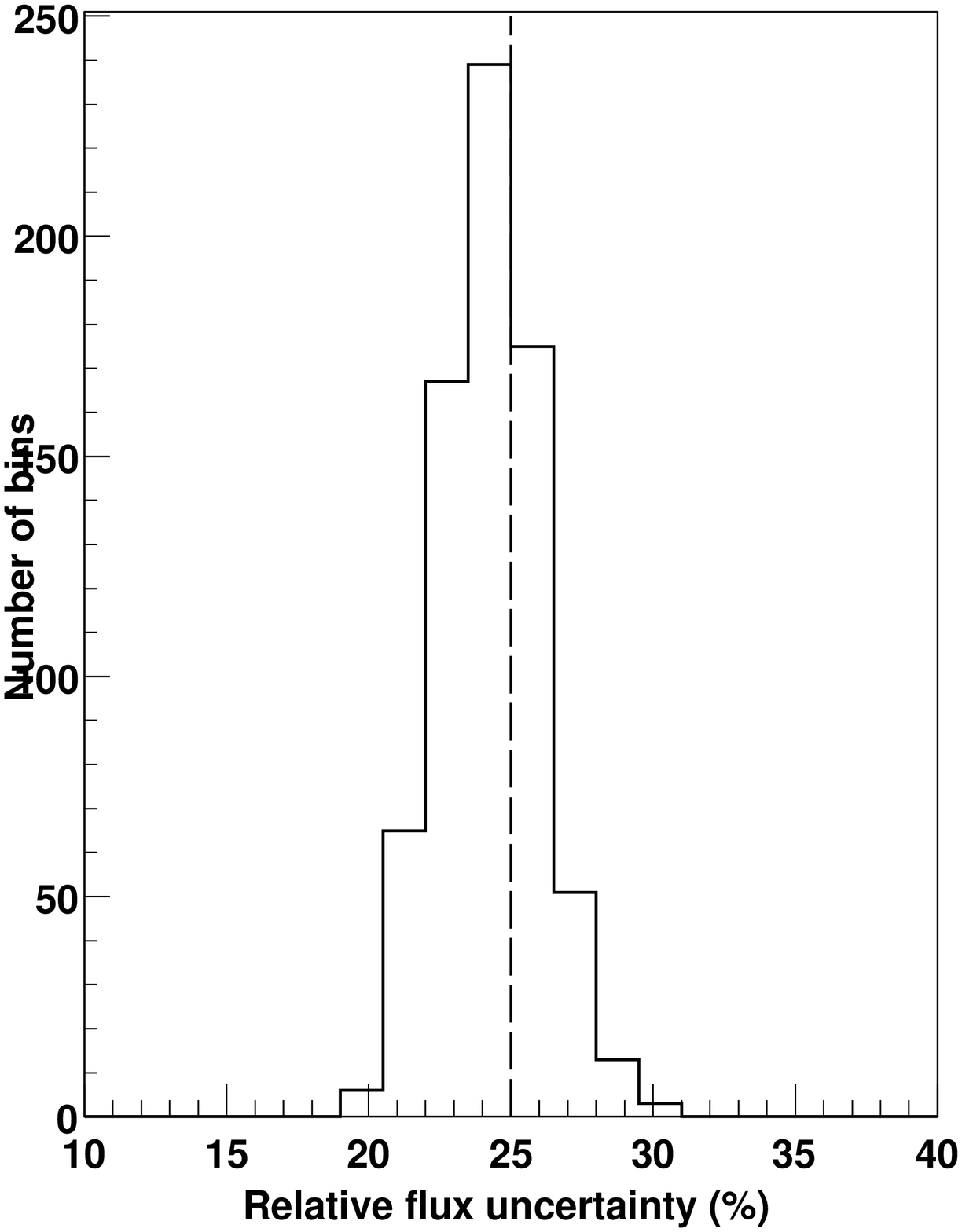}
 \includegraphics[scale=0.15]{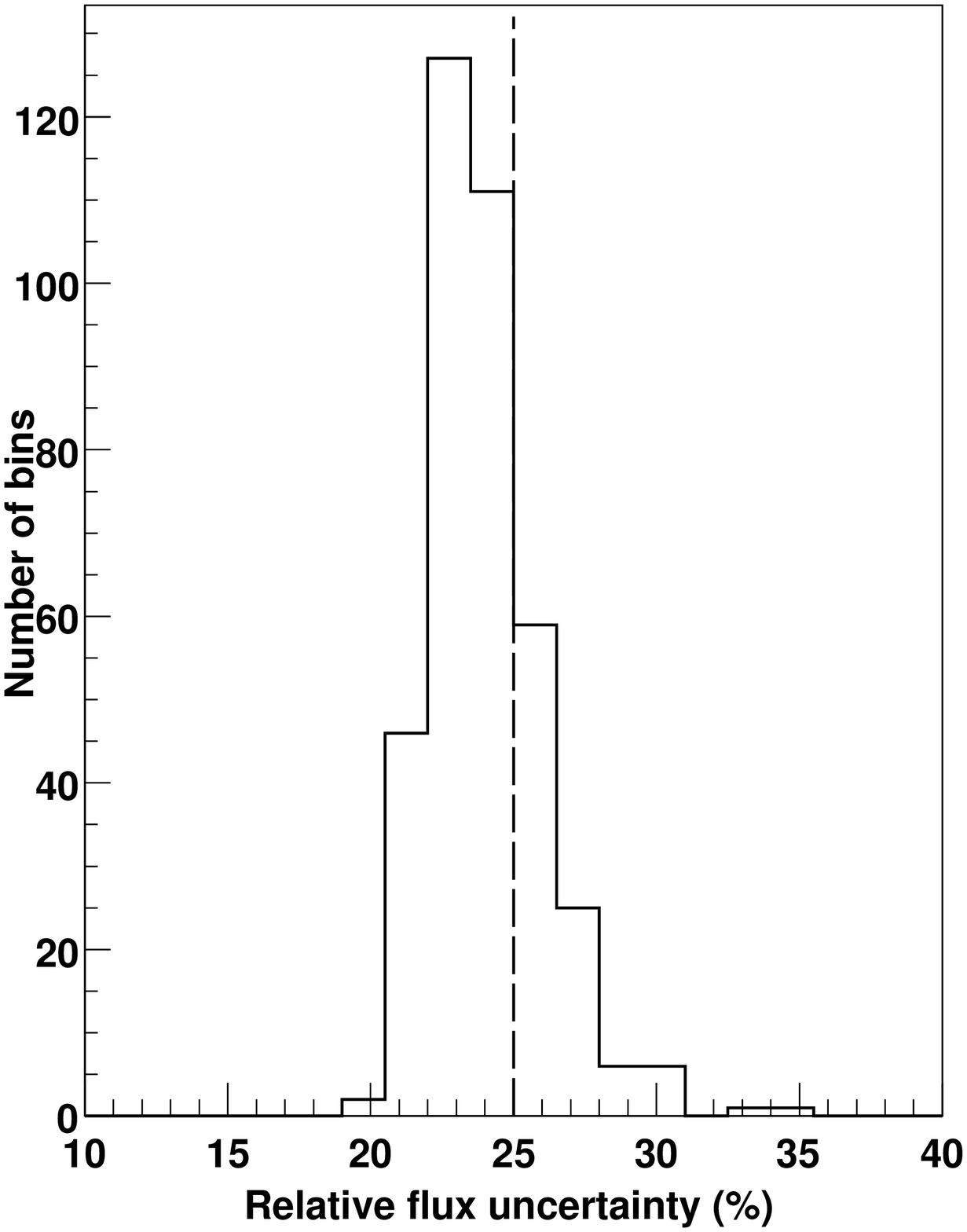}
  \caption{Distribution of $\sigma_{\ln\,F}$ for $\Delta_0$=25\% (depicted by a dashed line) for the different parameter sets: $\Gamma$=2.0 (left) and $\Gamma$=2.4 (right), $F_{100}$=5, 1, 0.5  from top to bottom.}

  \label{fig:uncfdist}

\end{figure}

The distributions of relative statistical uncertainties obtained with the adaptive-binning method (resulting from Step 2) are presented in  Figure \ref{fig:uncfdist}. These distributions are more gaussian-like than those obtained for variable sources, with a mean value close to 25\% as required and a typical rms of about 1.7\%. Simulations performed for a very low flux of $F_{100}$=0.1 ($\Gamma$=2.0), where the method yields a single bin over the 11-month period, show that the mean of the distribution remains within 10\% of the target value even in that extreme case.   

\begin{table}
\begin{centering} 
\begin{tabular}{|c|c|c|c|c|c|}
\hline
 & \multicolumn{2}{c|}{$\Gamma$ = 2.0} & \multicolumn{2}{c|}{$\Gamma$ = 2.4}\\
\hline
$F_{100}$  & $<N_{bins}>$ & $E_1$ (MeV)  & $<N_{bins}>$ & $E_1$ (MeV)\\
\hline
0.5 & 14 & 396 & 8 & 299 \\
1 & 37 & 332 & 23 & 253 \\
5 & 268 & 246 & 192 & 205\\
\hline
\end{tabular}
\caption{Average number of time bins and optimum energies for the different simulation conditions.\label{tab:npts}}
\end{centering}
\end{table}

The possible biases induced by this method are explored in the following.

\subsubsection{Bias and statistical fluctuations in flux and photon-index measurements}

The {\sl pylikelihood} analysis is expected to provide the correct mean flux and photon spectral index over all possible time intervals. We nevertheless check for the absence of biases in the measurement of these parameters when the time intervals are determined with the present method.  Furthermore, possible additional fluctuations in the measured parameters could arise from the intrinsic correlation between bin width and photon contents of the bin. Both effects are addressed here.

\begin{figure}[!h]

 \centering

 \includegraphics[scale=0.45]{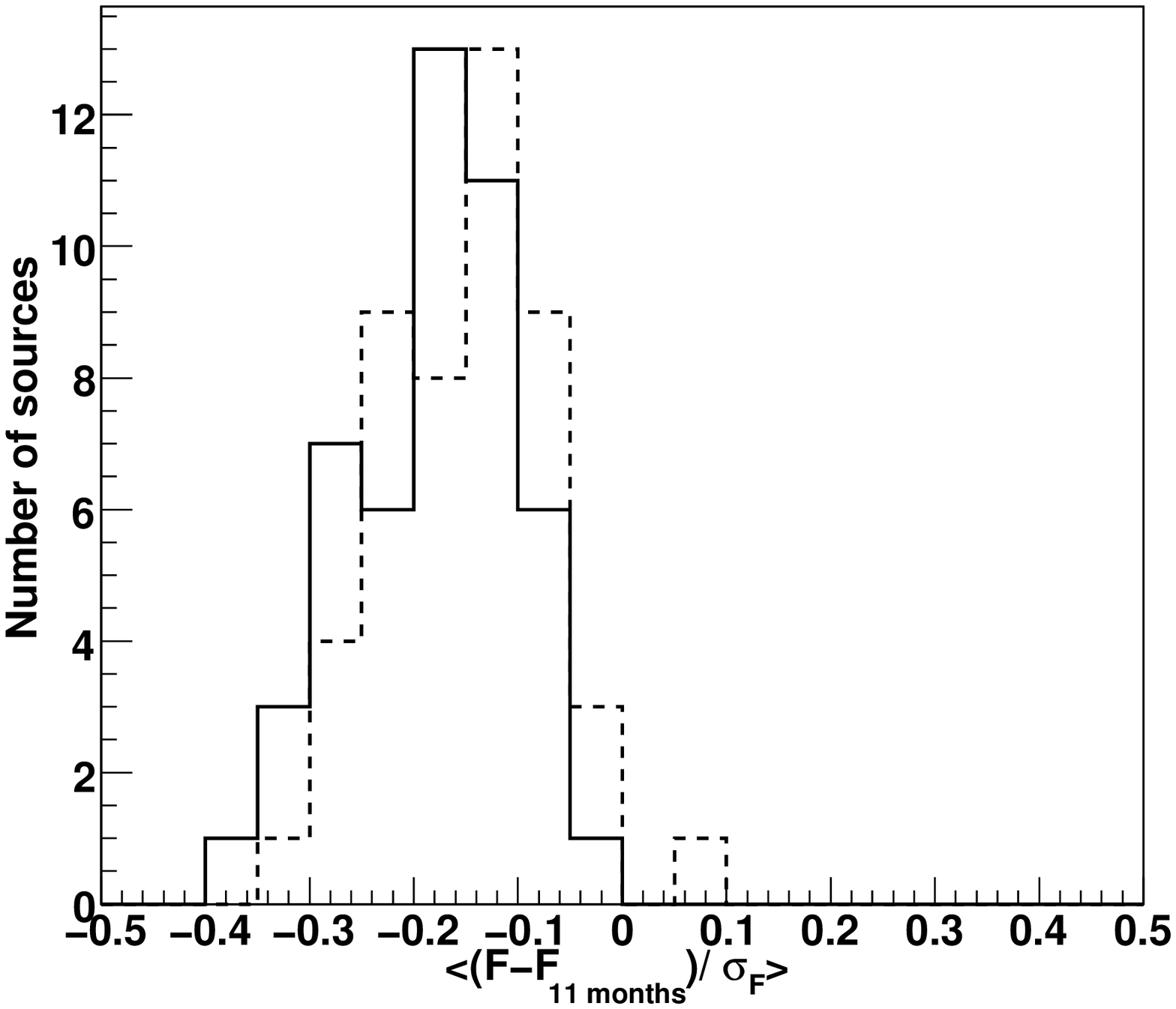}
 \includegraphics[scale=0.45]{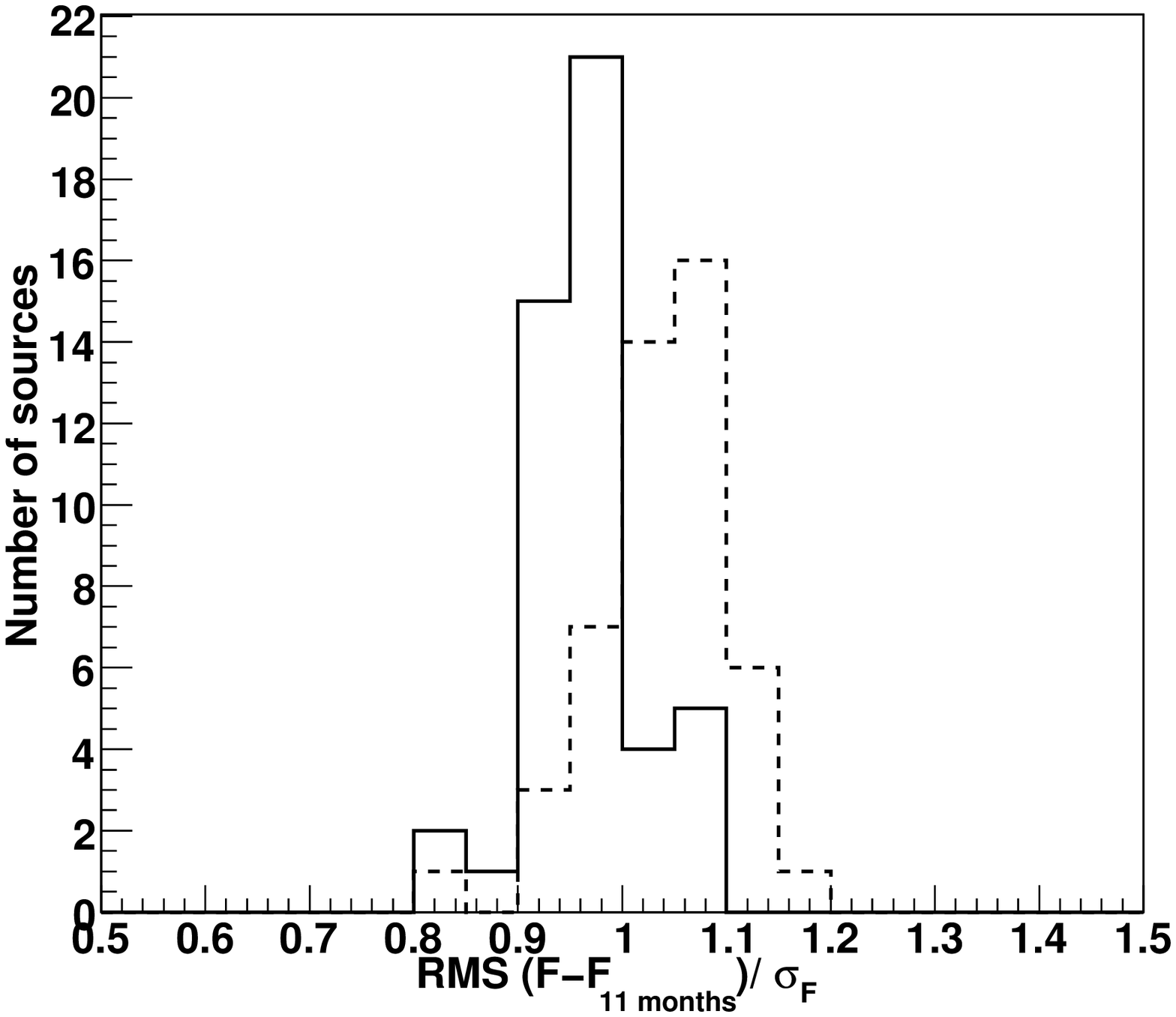} 
 \includegraphics[scale=0.45]{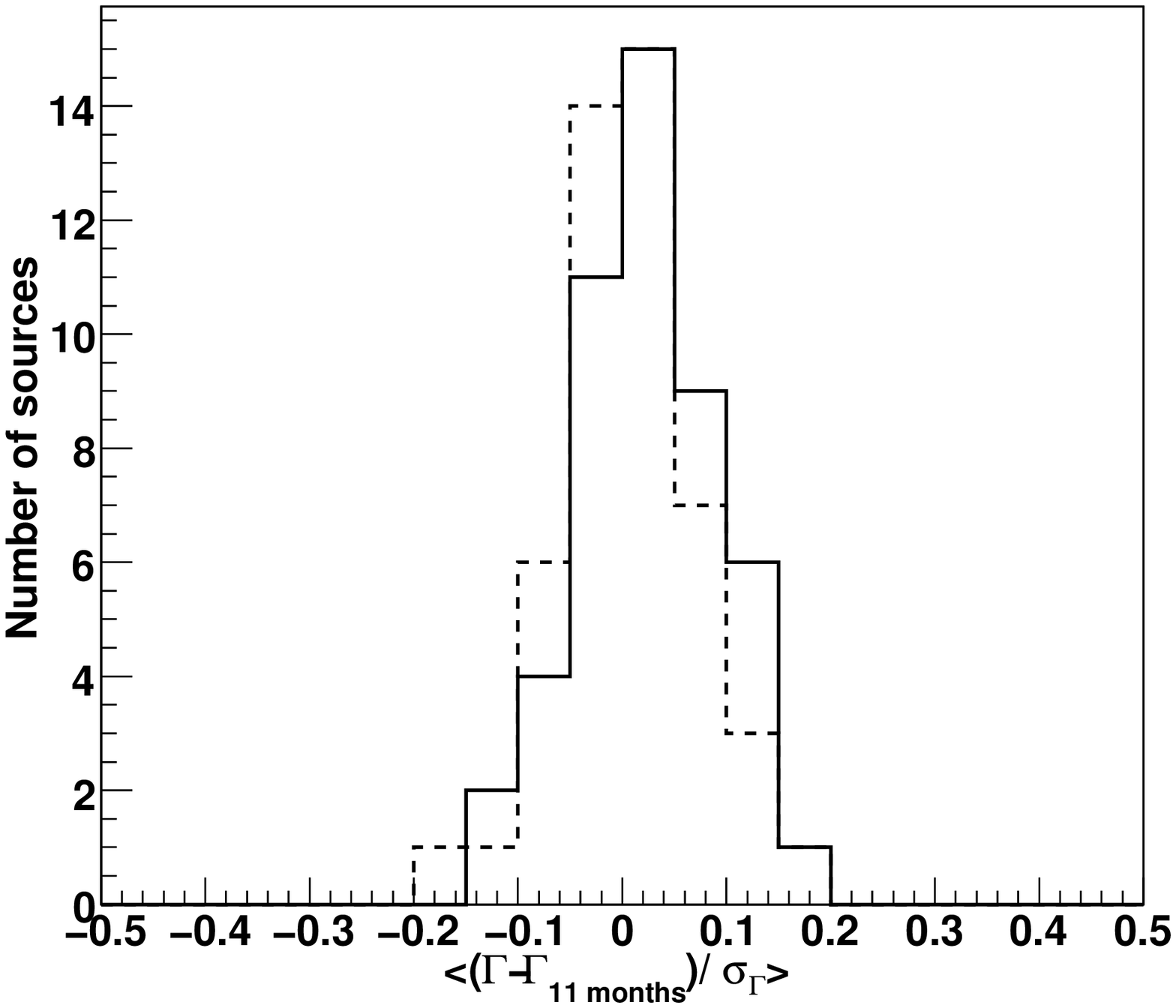}
 \includegraphics[scale=0.45]{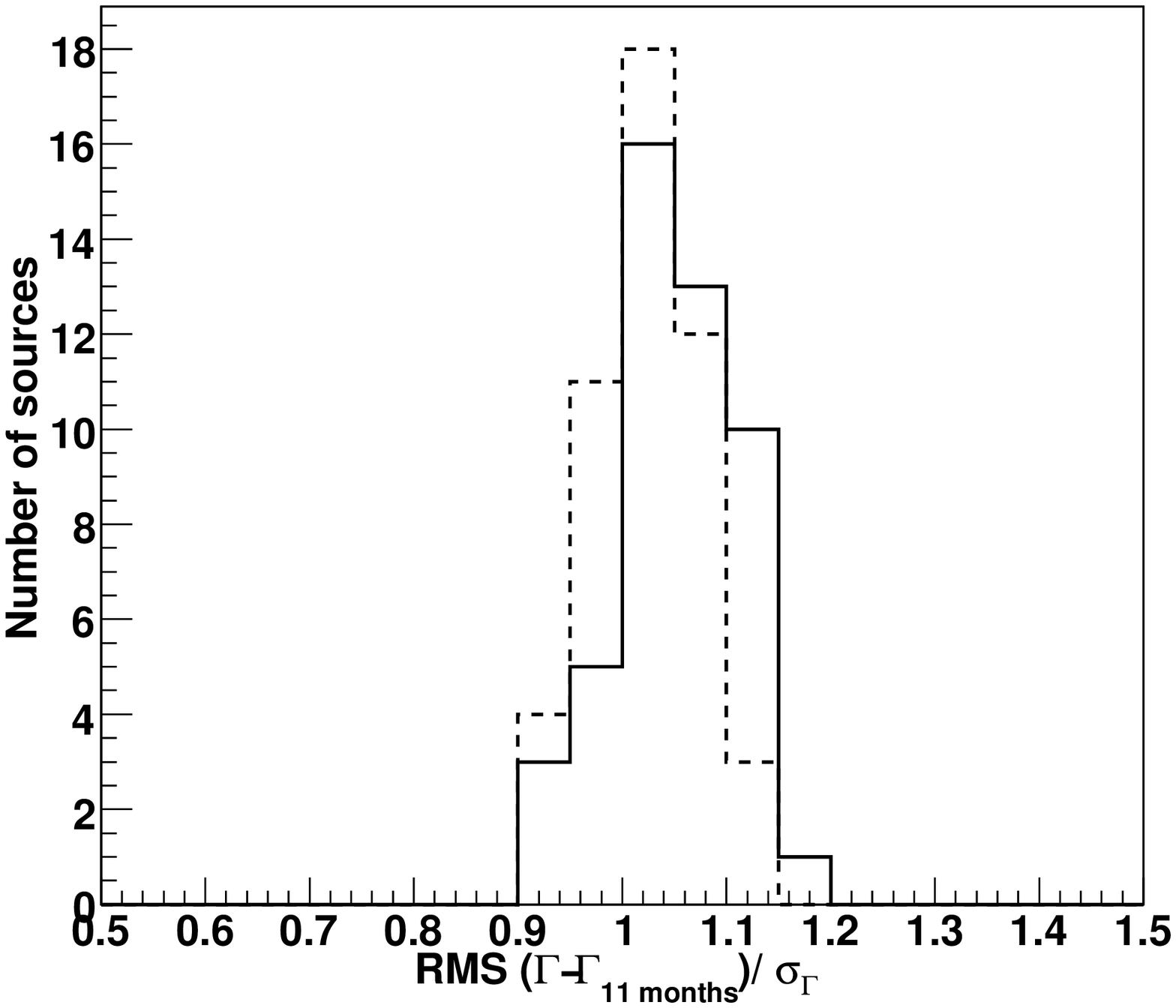} 
  \caption{Distributions of  mean values (left) and rms (right) for flux (top) and photon index (bottom) obtained in adaptive- (solid) and fixed-binning (dashed) light curves. In this example, $\Gamma$ = 2.4 and $F_{100}$=5. }
   \label{fig:mean_rms}

\end{figure}

One defines ${\Delta F}$  (${\Delta \Gamma}$) as  the difference, expressed in sigma,
between the measured flux (index) and the average value measured over the whole 11-month period.  
 Summing over all realizations, the distributions of the means (left) and rms (right) of the ${\Delta F}$ (top) and ${\Delta \Gamma}$ (bottom) distributions are displayed in Figure \ref{fig:mean_rms} for a particular set of parameters.  The results of the adaptive- (solid) and fixed- (dashed) binning methods are compared in this figure.  The  means $<{\Delta F}>$  and $<{\Delta \Gamma}>$ are listed 
in Tables \ref{tab:dist_mean_flux} and \ref{tab:dist_mean_index} respectively,  while the rms values are given in  Tables \ref{tab:dist_rms_flux} and \ref{tab:dist_rms_index}.

\begin{table}
\centering
\noindent \begin{tabular}{|c|c|c|c|c|}
\hline
$F_{100}$ & Adaptive ($\Gamma$=2.0) & Fixed ($\Gamma$=2.0) & Adaptive ($\Gamma$=2.4) & Fixed ($\Gamma$=2.4)\\
\hline
5 & -0.20$\pm$0.01 &  -0.15$\pm$0.01  & -0.19$\pm$0.01 & -0.15$\pm$0.01 \\
\hline
1 & -0.13$\pm$0.02 &  -0.09$\pm$0.02  & -0.14$\pm$0.03 & -0.13$\pm$0.03 \\
\hline
0.5 & -0.06$\pm$0.03 &  -0.07$\pm$0.03  & -0.12$\pm$0.05 & -0.13$\pm$0.05 \\
\hline
\end{tabular}
\caption{Comparison of $<{\Delta F}>$ for adaptive and fixed binnings.}
\label{tab:dist_mean_flux}
\end{table}
\begin{table}
\centering
\noindent \begin{tabular}{|c|c|c|c|c|}
\hline
$F_{100}$ & Adaptive ($\Gamma$=2.0) & Fixed ($\Gamma$=2.0) & Adaptive ($\Gamma$=2.4) & Fixed ($\Gamma$=2.4)\\
\hline
5 & 0.02$\pm$0.01 &  0.02$\pm$0.01  & 0.02$\pm$0.01 & 0.01$\pm$0.01 \\
\hline
1 & 0.02$\pm$0.03 &  0.05$\pm$0.03  & 0.04$\pm$0.03 & -0.01$\pm$0.03 \\
\hline
0.5 & 0.06$\pm$0.03 &  0.05$\pm$0.03  & 0.03$\pm$0.06 & 0.03$\pm$0.05 \\
\hline
\end{tabular}
\caption{Comparison of $<{\Delta \Gamma}>$ for adaptive and fixed binnings.}
\label{tab:dist_mean_index}
\end{table}
\begin{table}
\centering
\noindent \begin{tabular}{|c|c|c|c|c|}
\hline
$F_{100}$ & Adaptive ($\Gamma$=2.0) & Fixed ($\Gamma$=2.0) & Adaptive ($\Gamma$=2.4) & Fixed ($\Gamma$=2.4)\\
\hline
5 & 0.99$\pm$0.01 &  1.06$\pm$0.01  & 0.97$\pm$0.01 & 1.03$\pm$0.01 \\
\hline
1 & 0.96$\pm$0.02 &  1.01$\pm$0.01  & 0.93$\pm$0.02 & 0.98$\pm$0.02 \\
\hline
0.5 & 0.90$\pm$0.03 &  1.02$\pm$0.03  & 0.81$\pm$0.03 & 0.87$\pm$0.03 \\
\hline
\end{tabular}
\caption{Comparison of the average rms of the ${\Delta F}$  distributions for  adaptive and fixed binnings.}
\label{tab:dist_rms_flux}
\end{table}
\begin{table}
\centering
\noindent \begin{tabular}{|c|c|c|c|c|}
\hline
$F_{100}$ & Adaptive ($\Gamma$=2.0) & Fixed ($\Gamma$=2.0) & Adaptive ($\Gamma$=2.4) & Fixed ($\Gamma$=2.4)\\
\hline
5 & 1.02$\pm$0.01 &  1.03$\pm$0.01  & 1.05$\pm$0.01 & 1.03$\pm$0.01 \\
\hline
1 & 1.00$\pm$0.02 &  1.04$\pm$0.02  & 1.05$\pm$0.02 & 1.05$\pm$0.02 \\
\hline
0.5 & 1.00$\pm$0.03 &  1.03$\pm$0.02  & 1.05$\pm$0.04 & 1.01$\pm$0.05 \\
\hline
\end{tabular}
\caption{Comparison of the average rms of the ${\Delta \Gamma}$ distributions for adaptive and fixed binnings.}
\label{tab:dist_rms_index}
\end{table}
A small bias is observed in the mean flux values. Since similar features are seen for both adaptive- and fixed-binning light curves, one can safely conclude that this bias arises from small systematics effects pertaining to our {\sl pylikelihood} analysis and is not created by the adaptive-binning approach. 

The average rms values are found to be close to the expected value of 1 for all cases (with slightly lower values for the adaptive-binning approach relative to the fixed-binning one). One can thus exclude that the adaptive-binning approach gives rise to significant fluctuations in the measured parameters.

\subsubsection{Flux-index correlation}

\begin{figure}[ht!]

 \centering

 \includegraphics[scale=0.45]{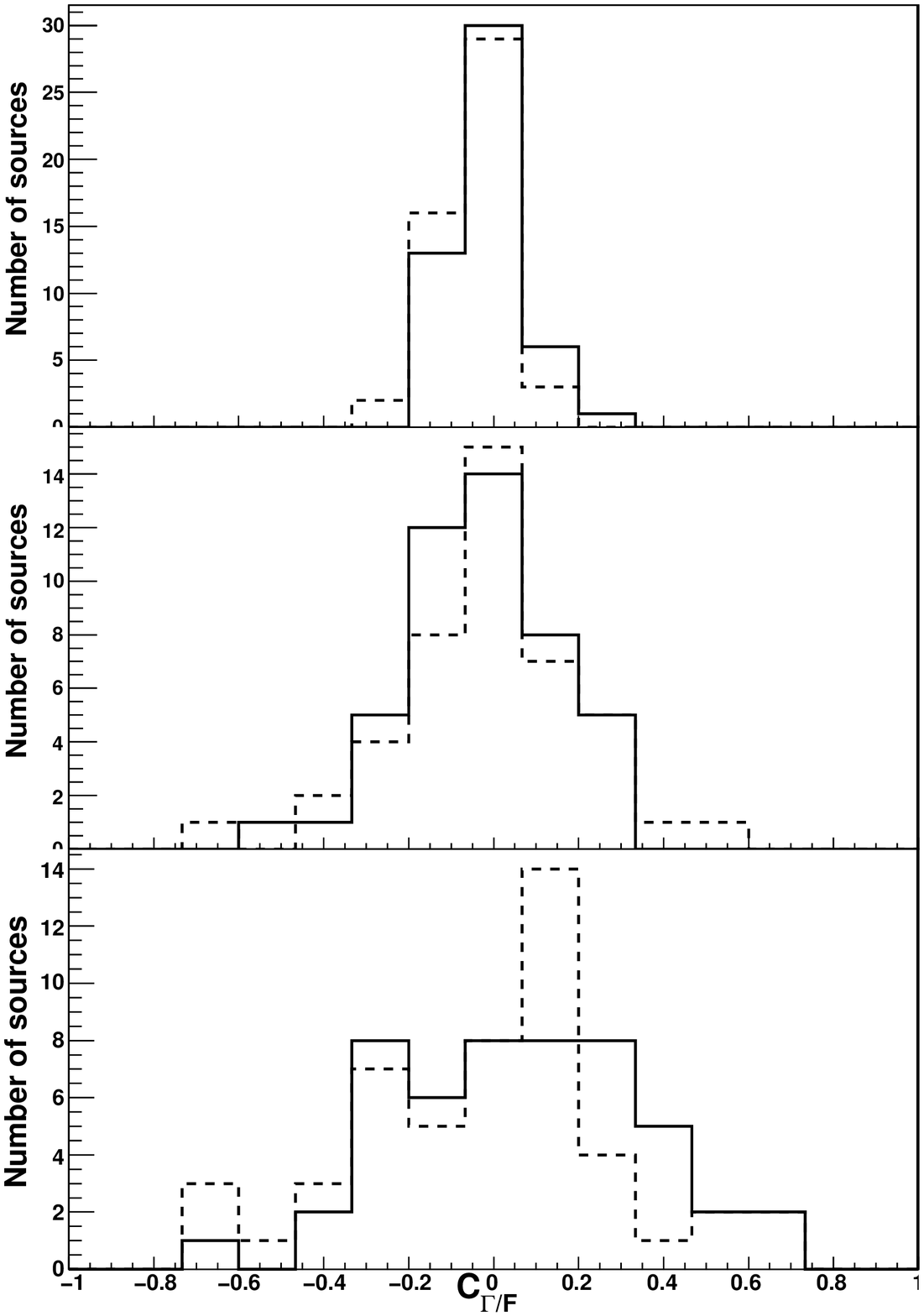}
 \includegraphics[scale=0.45]{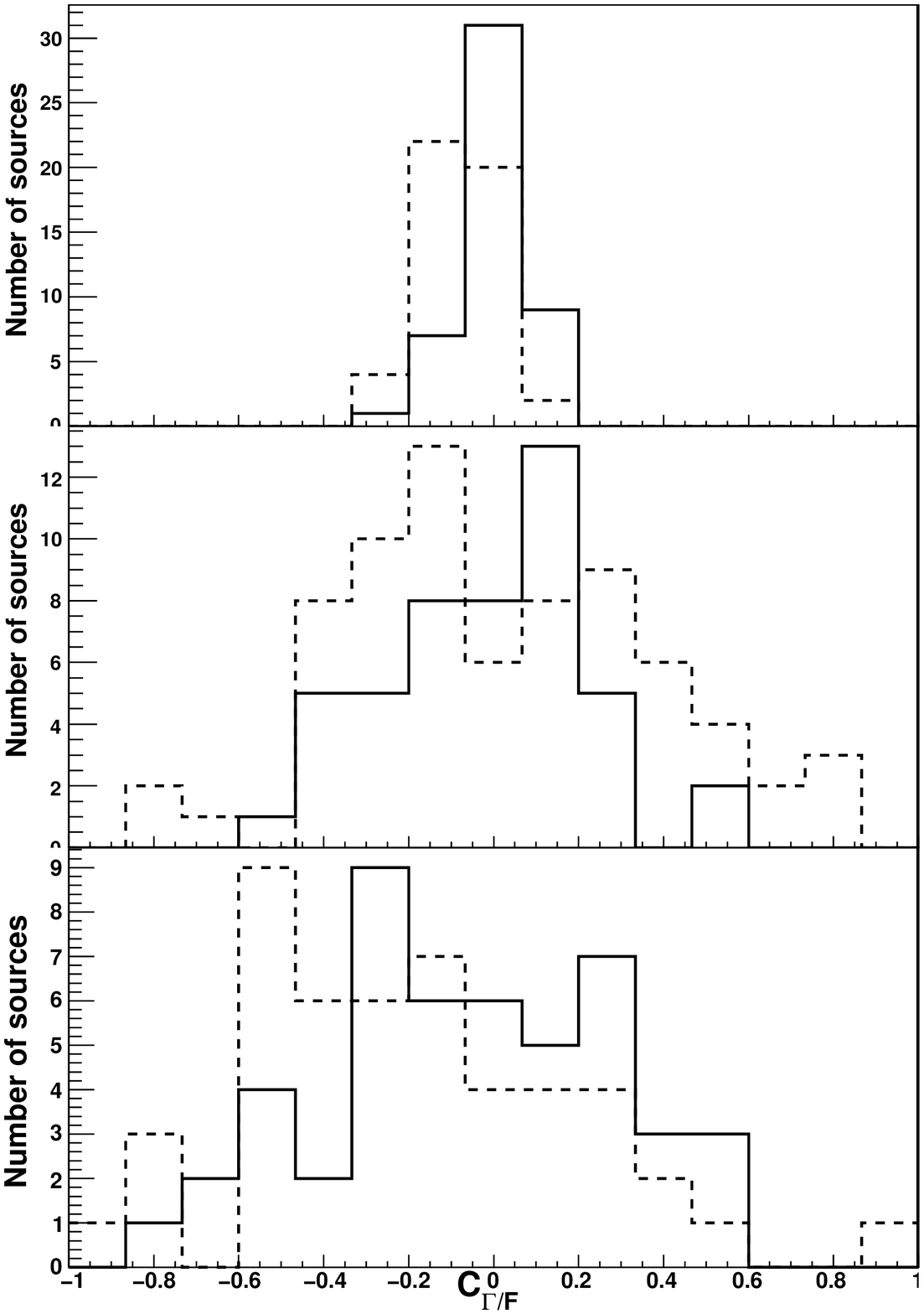}
  \caption{Distributions of correlation factor between spectral index and flux  for adaptive (solid) and fixed (dashed) binnings. $\Gamma$=2.0 (left), $\Gamma$=2.4 (right), $F_{100}$=5, 1, 0.5  from top to bottom.}
  \label{fig:cor_flux_ind_adfi}

\end{figure}

Another issue worth investigating is the correlation between measured photon index and flux in a given time interval. All photons do not contribute equally to the fulfillment of the criterion, with higher-energy photons contributing more. The detection of a series of photons with higher-than-average energies will thus lead to the criterion being satisfied faster and consequently to the interval under consideration being closed earlier.  In order to evaluate whether the adaptive-binning method creates additional correlation between the photon spectral index and flux (as measured in Step 2), the corresponding correlation factor is computed for each realization.  Since $E_{min}$ is chosen as the optimum energy, for which  the correlation between photon index and integral flux is minimum by definition, the observed correlation factor is expected to be small.  
The correlation-factor distributions obtained for adaptive and fixed binnings are shown in Figure \ref{fig:cor_flux_ind_adfi}.
For all cases, the distributions are very similar for the two approaches and are all centered at small values, with larger fluctuations for lower fluxes. Therefore, the measured photon spectral indices and fluxes do not appear more correlated when using an adaptive binning relative to a fixed binning.

\subsubsection{Interbin correlation}

\begin{figure}[ht!]

 \centering

 \includegraphics[scale=0.3]{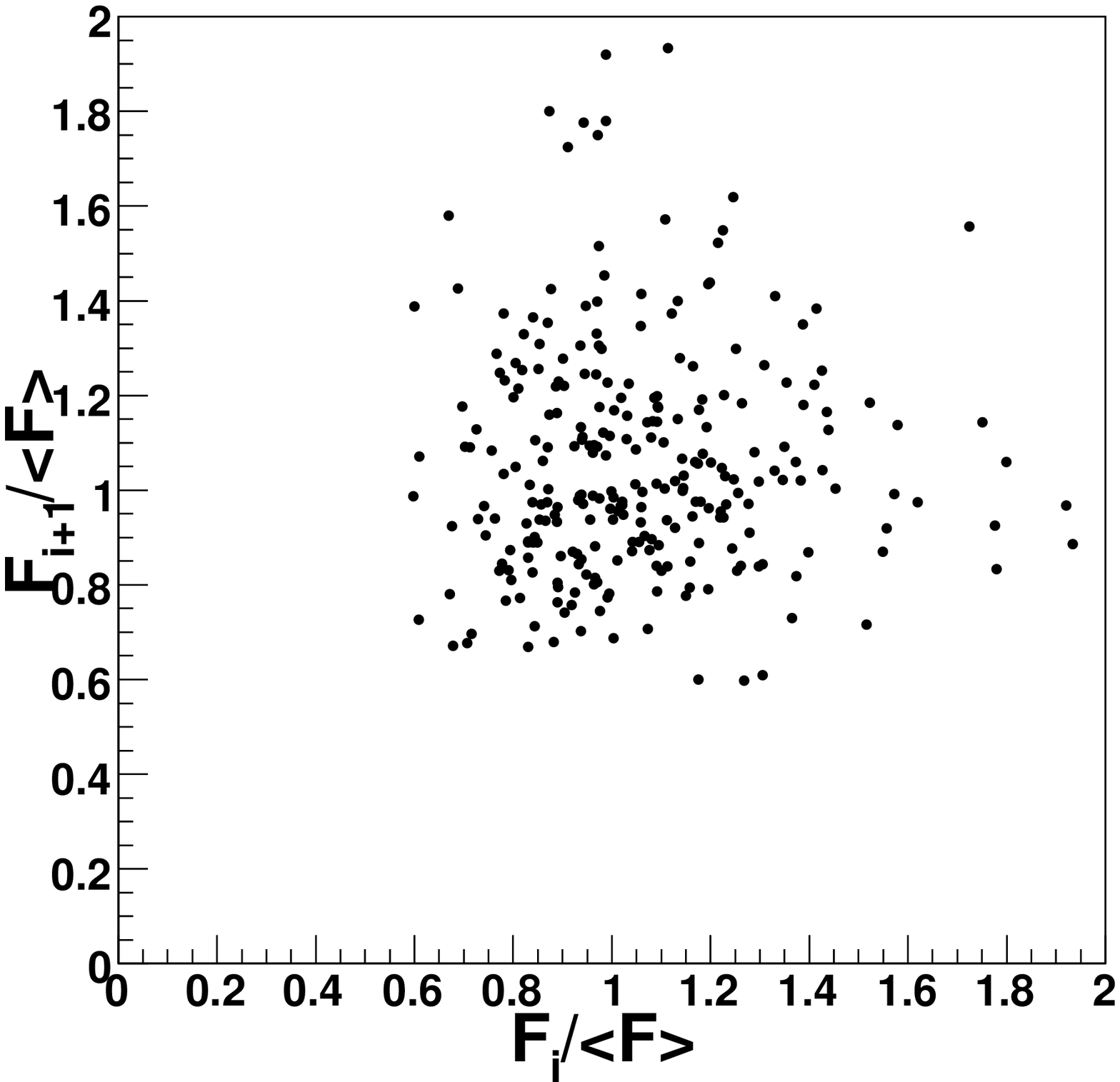}
 \includegraphics[scale=0.3]{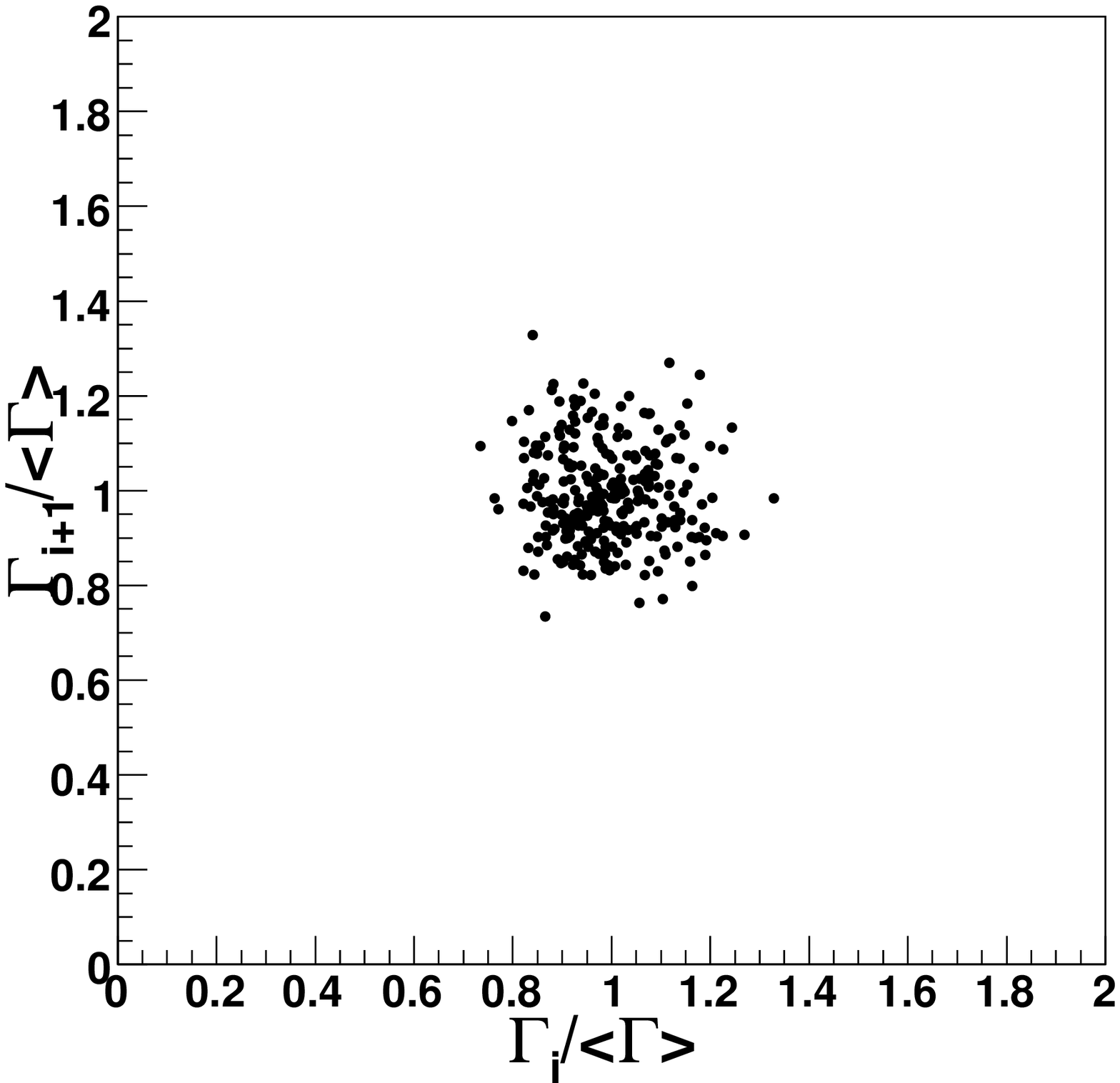} 
  \caption{Fluxes  (left) and photon spectral indices (right) measured in consecutive bins plotted one against another for a given realization.}

  \label{fig:corr_ex}

\end{figure}

In the adaptive-binning method, the starting time of a given bin depends on the starting time and flux-dependent width of the previous bin and thus by recurrence on the whole flux history since the start of the light curve. In this section, we explore the impact of the interbin correlation
on the measured source parameters by comparing these in two consecutive bins. A comparison will again be made with the results of a similar analysis performed with a fixed binning,  for which fluxes measured in consecutive bins are independent by nature.  The flux (index) in one bin  is plotted versus 
the flux (index) in the next bin in Figure \ref{fig:corr_ex} left (right) for the whole light curve of a particular realization.

\begin{figure}[ht!]

 \centering

 \includegraphics[scale=0.4]{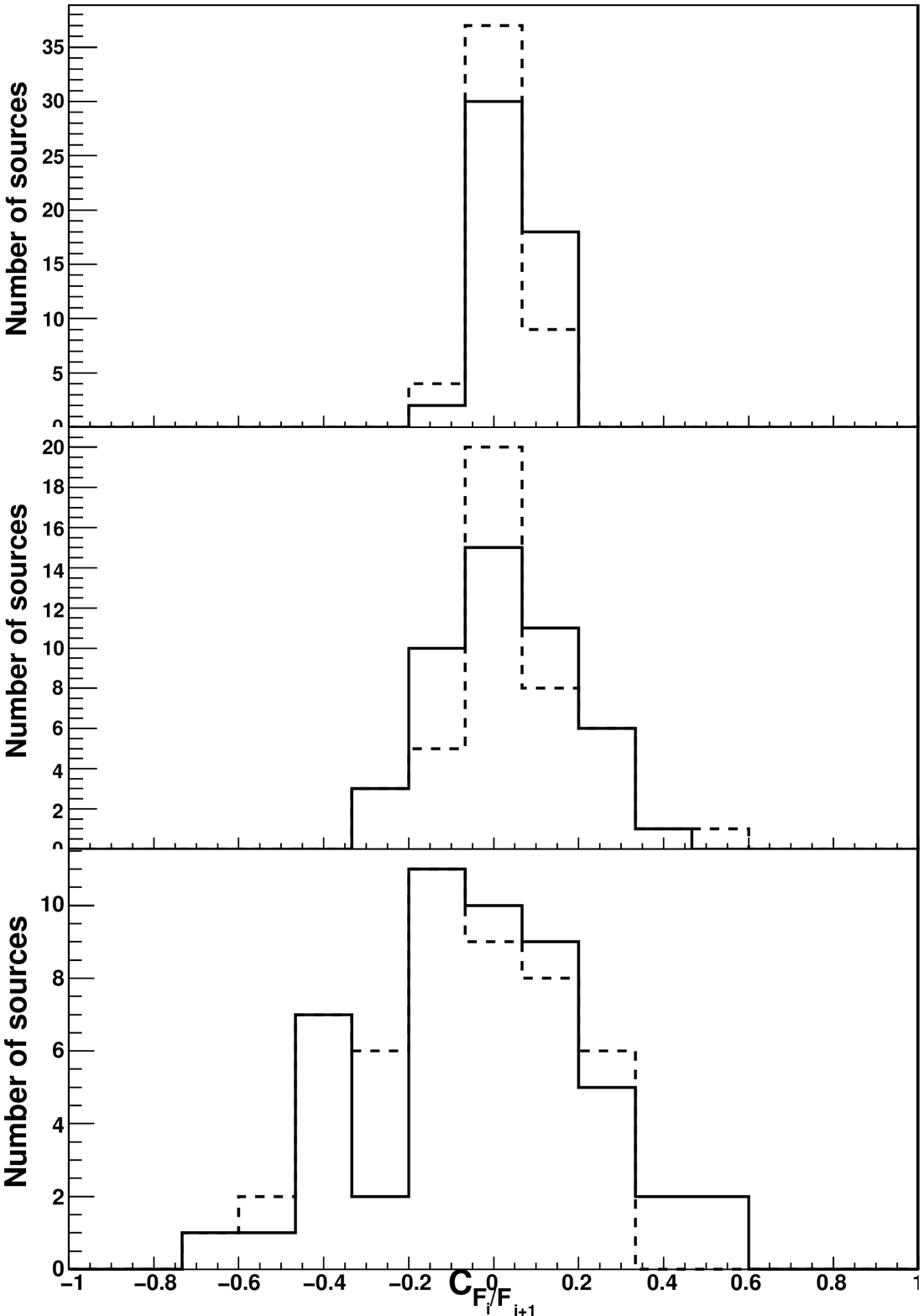}
 \includegraphics[scale=0.4]{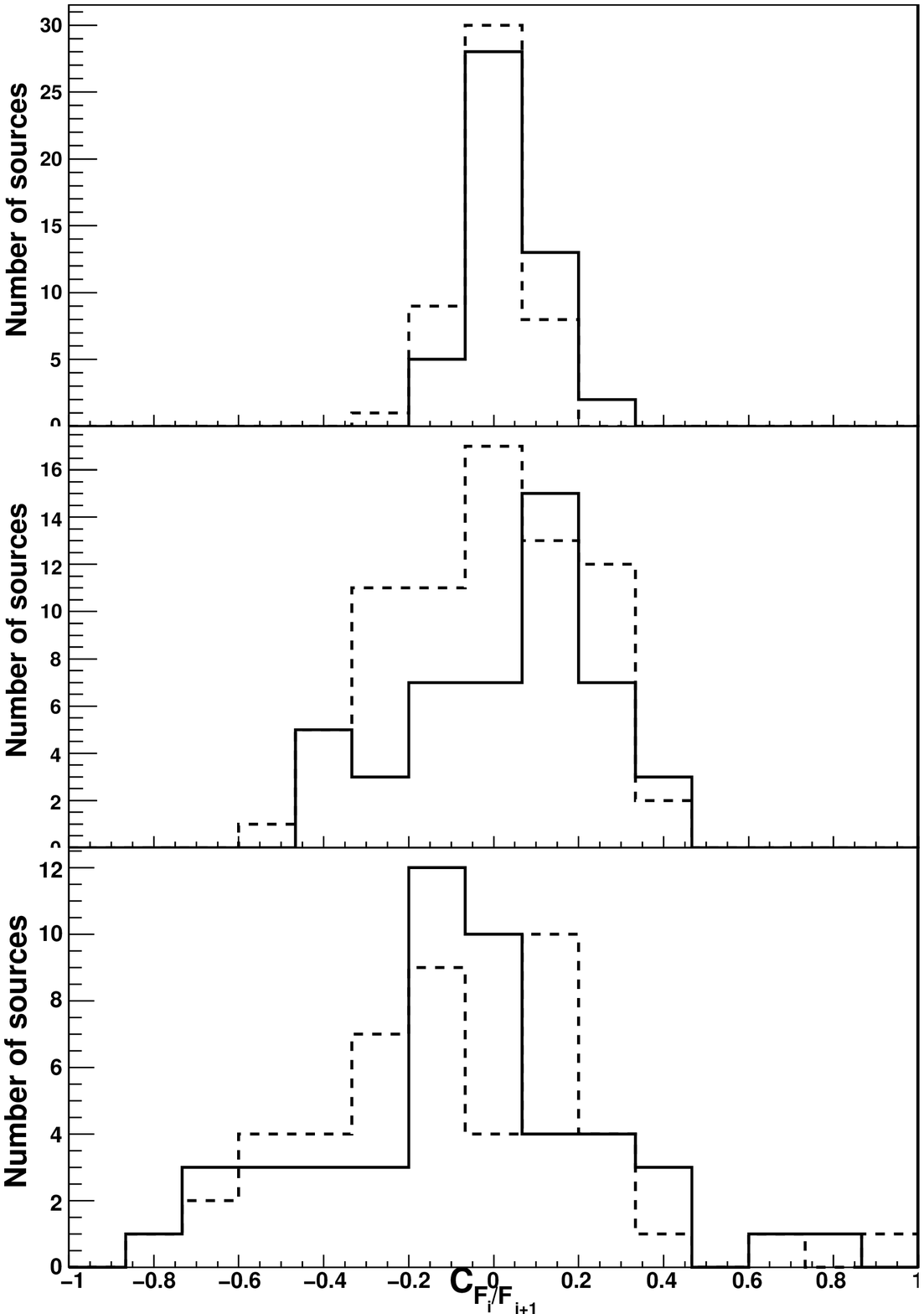}
 \includegraphics[scale=0.4]{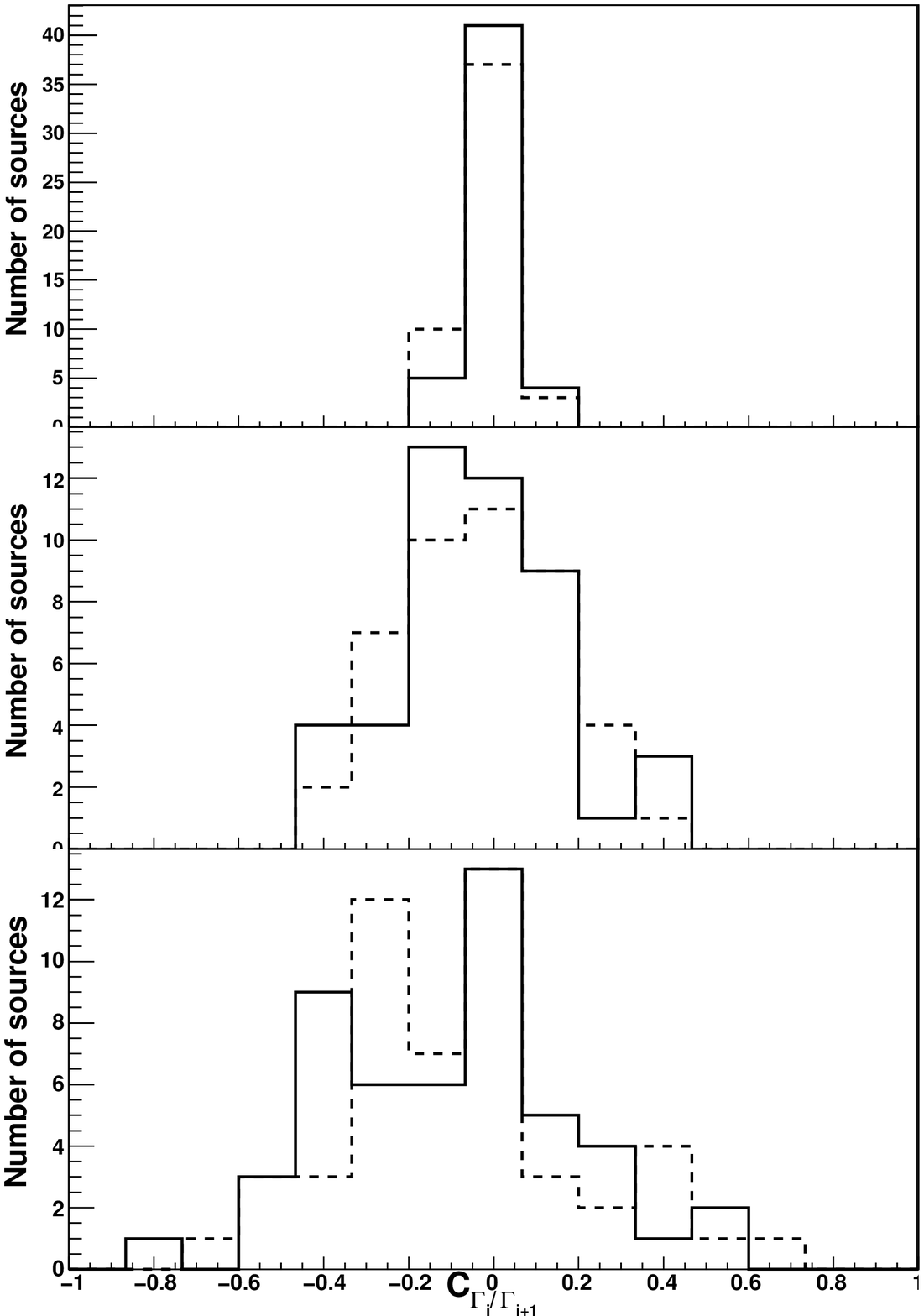}
 \includegraphics[scale=0.4]{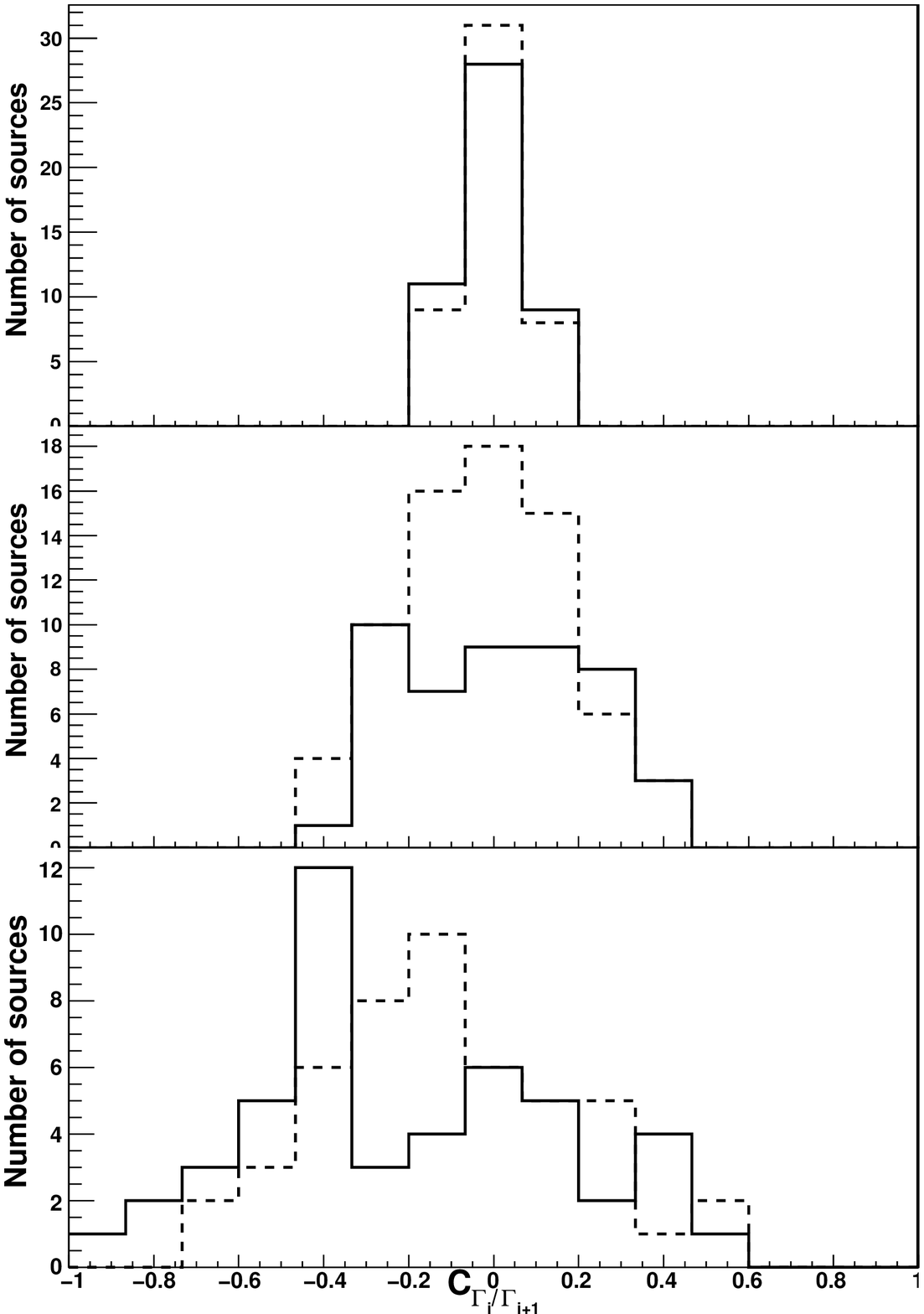}
  \caption{Distributions of correlation factors between consecutive bins both for flux (top panels) and photon spectral index (bottom panels). 
  $\Gamma$=2.0 (left-hand panels) and $\Gamma$=2.4 (right-hand panels), $F_{100}$=5, 1, 0.5  from top to bottom in each panel. Solid: adaptive binning, dashed: fixed binning}

  \label{fig:cor_consecutive_adfi}

\end{figure}

The distribution of correlation factors obtained with the different realizations are presented in Figure \ref{fig:cor_consecutive_adfi}. No significant differences are observed between distributions obtained with adaptive and fixed binnings for all cases. Both distributions are centered close to 0 (as expected for the fixed-binning case), demonstrating that the interbin dependence intrinsic to the adaptive binning method has little impact on the correlation between parameters measured in consecutive bins.

\section{High-level analysis with variable sources}

\subsection{Duty cycle}
\begin{figure}[!ht]

\centering
\includegraphics[scale=1]{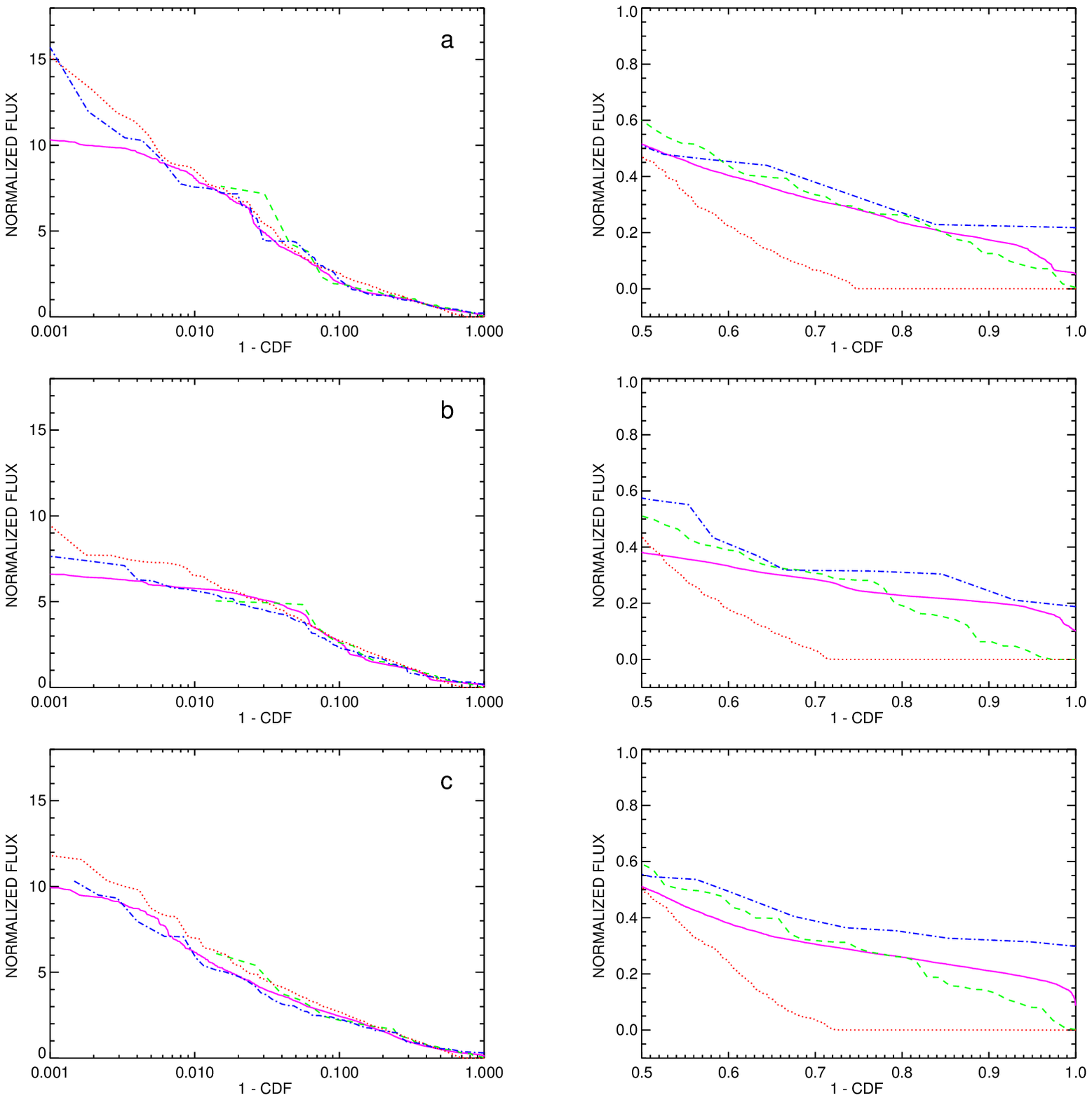} 
\caption{The normalized flux is plotted against the parameter 1-CDF, obtained for the different light curves shown in Fig. \ref{fig:lc_var}. Left: whole range, right: zoom on the low-CDF region. Solid purple: true light curve, dashed green: fixed binning, dotted red: fixed binning with higher resolution, dot-dashed blue: adaptive binning.}

\label{fig:dc}

\end{figure}

One way to describe variations in a light curve is via a duty cycle, corresponding to the fraction of time a source is
in a ``bright'' state. More generally this can be presented as a flux distribution
or its integral, the cumulative distribution function (CDF). In Figure \ref{fig:dc} we have
plotted flux, normalized to a mean of 1, versus 1-CDF, for the light curves
displayed in Figure \ref{fig:lc_var}. The curves give the fraction of time (1-CDF) that
a source is above a particular flux level. Four curves are shown for
each light curve, the `true' light curve, the adaptive-binning light curve, the fixed-binning light curve (with the same numbers of bins as the adaptive-binning light curve)  and the fixed-binning light curve with higher (16 times) time resolution. The two
fixed-binning light curves are complementary in that the high
resolution one works well at the high-flux end but fails at the
faint-end where the noise governs the distribution because of the low
signal-to-noise ratio of the individual points.
With adaptive binning the bin widths can be optimized such that in the
presence of noise, the flux distribution, or duty cycle, can be described
over a much wider range than is possible when a fixed bin width is used.
The larger the range in fluxes the more advantageous
the adaptive binning is for this type of analysis. 

\subsection{Power Density Spectrum (PDS)}

\begin{figure}[ht]

\centering
\vspace{6cm}
\includegraphics[scale=0.5]{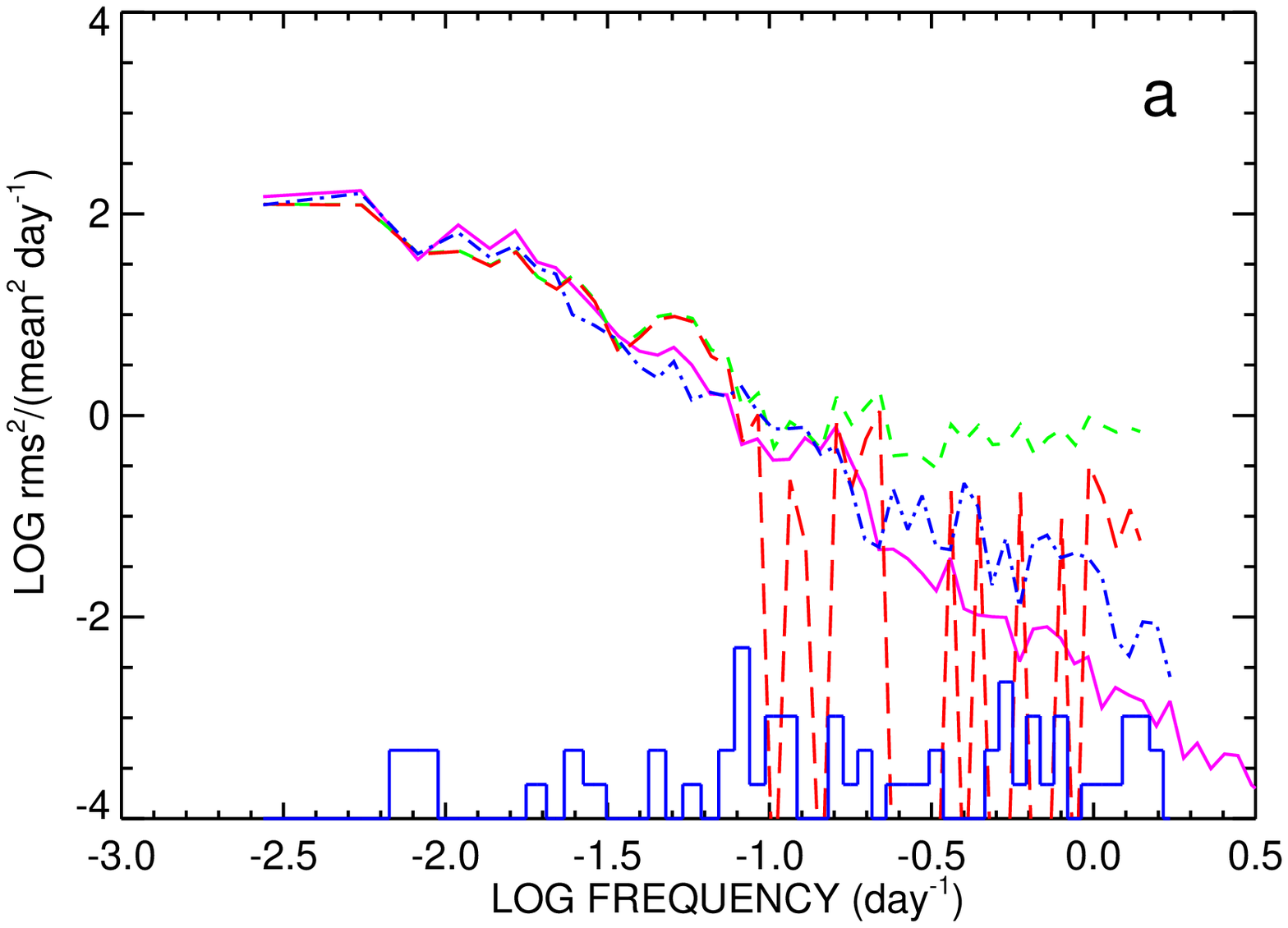} \\
\vspace{6cm}
\includegraphics[scale=0.5]{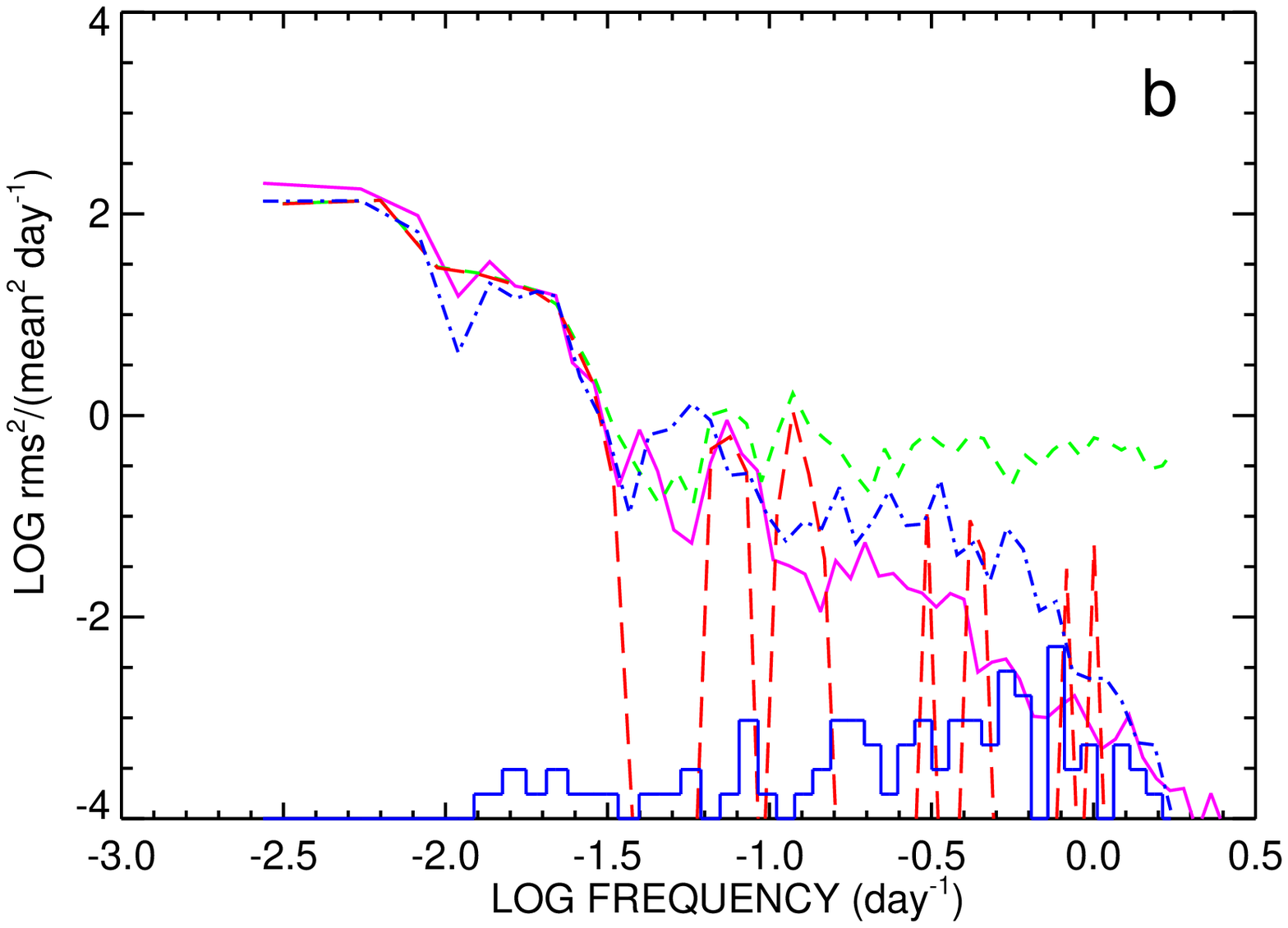} \\ 
\vspace{6cm}
\includegraphics[scale=0.5]{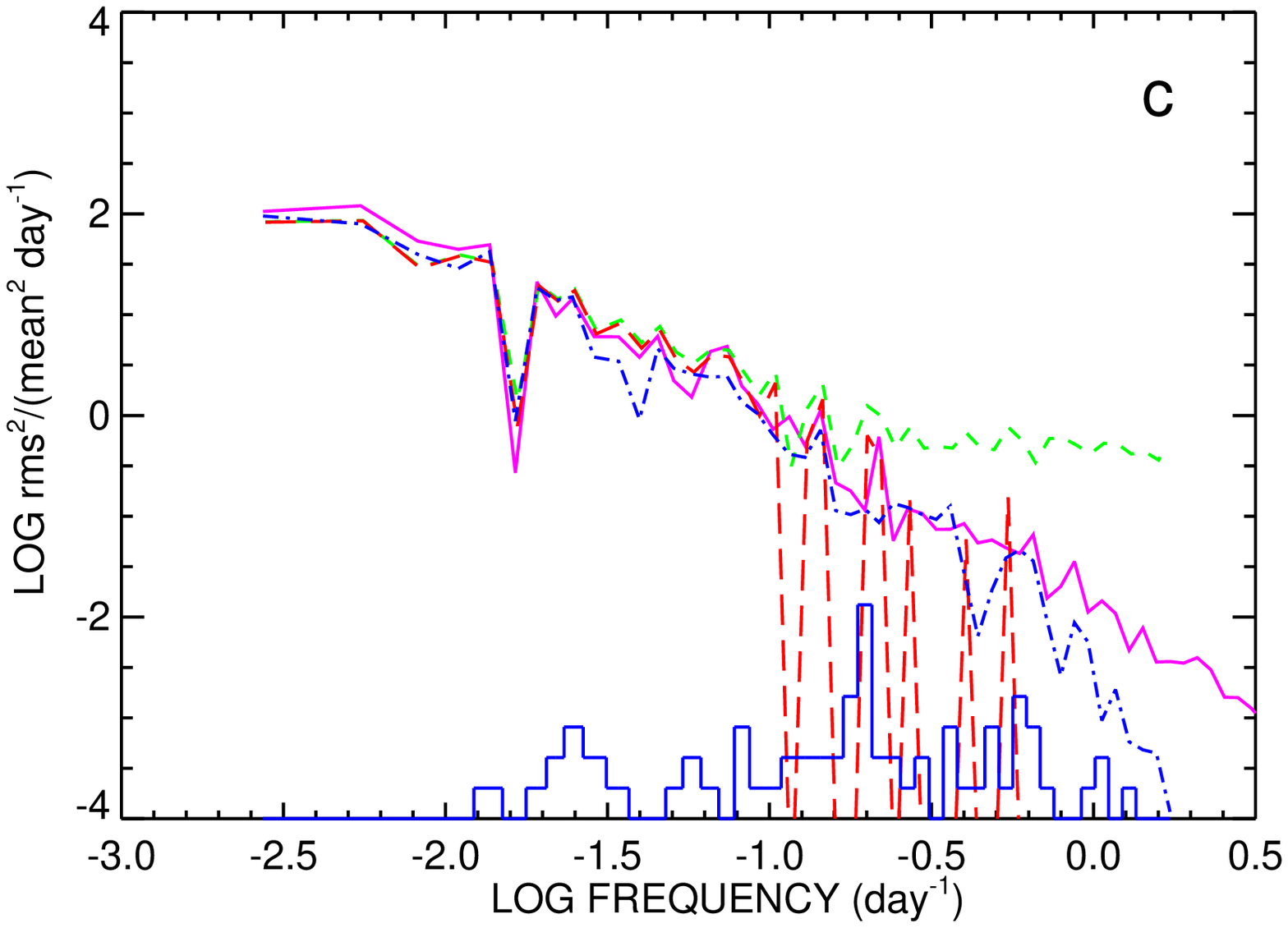}

\caption{Comparison of Power Density Spectra for the same simulations as in Fig. \ref{fig:lc_var}. In each plot the solid curve (purple) is the PDS of the
true light curve. The dashed (green) and long dashed (red)
curves are the PDS calculated for the fixed binned light curves, the white noise level being subtracted for the latter. The PDS of the adaptive binned light curve is
plotted as a dash-dot curve. All PDS curves were smoothed by averaging over logarithmic frequency bins. The distribution of
Nyquist frequencies corresponding to the time widths of
the adaptive bins is shown by the histogram (blue).
Note that this distribution is plotted in linear
scale, the zero level corresponding to the lower limit of the vertical axis.}
\label{fig:PDS}

\end{figure}

The PDS is one of the main tools used
to characterize light curve variability. For evenly sampled data the
PDS is computed directly via the Fourier transform of the light curve.
For light curves with uneven sampling the Lomb-Scargle
algorithm can be used and the PDS distortions caused by the
sampling are usually modeled by comparison with simulated
light curves \citep{Don92,Utt02}.
This method cannot easily be applied to adaptively binned
light curves since the bin widths depend on flux, making time sampling
different for each simulation. It is still possible to
make a qualitative estimate of the PDS for an adaptive-binning
light curve. To do this we oversample the light curve with
a linear interpolation between the center of each of the adaptive
time bins. The PDS is then computed with a standard Fourier transform.
The binning and interpolation modify the power density around and above
the frequencies corresponding to the widths of the time bins.
To check at what frequencies this becomes important for a particular
light curve we compute the distribution of Nyquist frequencies (equal to $0.5/dt$ where $dt$ is the bin width) associated with the bin widths. In Figure \ref{fig:PDS} this distribution
is plotted for the three simulated light curves shown in Figure \ref{fig:lc_var}, in each case together with the
PDS for the adaptive- and fixed-binning light curves as well as for the
{\it true} light curve. For the fixed-binning case we also show
the PDS after subtraction of the white noise level, illustrating
where, for this PDS, the signal is lost in the noise.
It is clear from these examples that the PDS of the adaptive-binning
light curve provides a qualitatively reasonable estimate
of the PDS over a similar frequency range as the PDS for the fixed
binning case. At the higher frequencies however, the adaptive-binning PDS,
as we compute it here, is determined essentially by the binning interpolation.
A more detailed discussion of the PDS properties and ways in which the binning effects can be taken into account is outside the scope of the present paper.

\section{Summary}

We have presented a method  enabling constant-uncertainty/constant-significance light curves to be computed with the {\sl Fermi}-LAT data. These light curves encapsulate more information than do fixed-binning light curves, without favoring any a priori arbitrary timescale and while avoiding upper limits.  Although primarily developed for blazar studies, the method is applicable to all variable sources. The current implementation allows the time intervals to be calculated in a simple and quick way, while the reported parameters result from the standard LAT analysis.  A slight skew in the measured flux distribution is a feature of the method. Possible adverse effects were explored by Monte-Carlo simulations but no significant issues have been revealed. This method opens capabilities regarding timing analysis in the $\gamma$-ray domain that were only available so far at longer wavelengths.
\section{Acknowledgments}
%\acknowledgments 

It is a pleasure to thank S. Fegan and J. Scargle for useful discussions and suggestions.

The \textit{Fermi} LAT Collaboration acknowledges generous ongoing support from a number of agencies and institutes that have supported both the development and the operation of the LAT as well as scientific data analysis.  These include the National Aeronautics and Space Administration and the Department of Energy in the United States, the Commissariat \`a l'Energie Atomique and the Centre National de la Recherche Scientifique / Institut National de Physique Nucl\'eaire et de Physique des Particules in France, the Agenzia Spaziale Italiana and the Istituto Nazionale di Fisica Nucleare in Italy, the Ministry of Education, Culture, Sports, Science and Technology (MEXT), High Energy Accelerator Research Organization (KEK) and Japan Aerospace Exploration Agency (JAXA) in Japan, and the K.~A.~Wallenberg Foundation, the Swedish Research Council and the Swedish National Space Board in Sweden. Additional support for science analysis during the operations phase is gratefully acknowledged from the Istituto Nazionale di Astrofisica in Italy and the Centre National d'\'Etudes Spatiales in France.  

\clearpage

\bibliography{adaptive.bib}

\clearpage

\begin{appendix}
\section{}
The goal is to estimate the relative integral-flux uncertainty $\sigma_{\ln\,F}$  from the data given particular source and background models. 
Both flux and $\sigma_{\ln\,F}$ estimates are based on an unbinned maximum-likelihood method presented in the following.
These estimates have been found to be in agreement with those ones given by the {\sl pylikelihood} analysis within a few percent. Some of the formulae derived below can also be found in the appendix of \cite{1fgl}.

\subsection{Log Likelihood}

The likelihood $\mathcal{L}$ is the probability of obtaining the data given an input model which is in this case the distribution of the $\gamma$-ray fluxes over the sky.
The maximum likelihood method consists in maximizing $\mathcal{L}$ over the parameter phase space of the model to get the best match between the model and the data. Following \cite{Cas79} (Eq. 7),
 the logarithm of the likelihood is obtained by adding up the contribution of each photon, with index i in the following, within a region of interest (ROI) over a time interval 
$[T_0,T_1]$:
\begin{eqnarray}
\ln \mathcal{L} =\sum\limits_i \ln M_i -N_{\rm pred}
\label{eq:loglike3}
\end{eqnarray} 
where $M_i=Exp(E_i,T_0,T_1)[S_s(E_i)PSF(r_i,E_i)+S_B(E_i)]$ is the model event density, $E_i$ is the photon energy, $r_i$ is the angular deviation from the source position,
 $Exp(E_i,T_0,T_1)$ is the exposure (time $\times$ effective area) over the considered time period,  $S_s(E_i)$ is the differential source spectrum, assumed to be a power law in the following, 
\begin{eqnarray}
S_s(E_i)=A \left( \frac{E_i}{E_0} \right)^{-\Gamma}
\label{eq:ss}
\end{eqnarray}
where $E_0$ is a reference energy, A is the differential flux for $E_i=E_0$ and $\Gamma$ is the photon spectral index, $PSF(r_i,E_i)$ is the point spread function\footnote{assumed to be independent of $\Theta_i$, the angle between the satellite axis and the arrival direction of the photons. This is justified for the LAT.}, 
$S_B(E_i)$ is the differential background per solid angle unit. Finally,
\begin{eqnarray} 
N_{\rm pred}=\int Exp(E,T_0,T_1) \left[ S_s(E)+\int S_B(E) d\Omega \right] dE
\label{eq:npred}
\end{eqnarray} 
is the number of photons within the ROI predicted by the model.

\subsection{Significance/Flux calculation}

The difference between $ln \mathcal{L}$ assuming that the source is present or absent is:
\begin{eqnarray}
\Delta \ln \mathcal{L}=\sum\limits_i \ln\left[1+g(r_i,E_i)\right] - N_{\rm src}
\label{eq:dlogl}
\end{eqnarray}
where $g(r_i,E_i)$ is the local source/background ratio
\begin{eqnarray}
g(r_i,E_i)= \frac{S_s(E_i)PSF(r_i,E_i)}{S_B(E_i)}
\label{eq:gi}
\end{eqnarray}
and $N_{\rm src}$ is the predicted number of photons coming from the source only:
\begin{eqnarray} 
N_{\rm src}=\int Exp(E,T_0,T_1)S_S(E)dE
\label{eq:nsrc}
\end{eqnarray} 
The flux of the source is estimated by maximizing $\Delta \ln \mathcal{L} $ with respect to A ($\Gamma$ is usually set fixed in Step 1).

The Test Statistic ($TS$) is defined as $2 \Delta \ln \mathcal{L}$. In principle, $TS$ behaves as a $\chi^2$ distribution with a number of degrees of freedom equal to the number of free source  parameters, but in our particular case the statistic is modified as negative flux values are not allowed in the fit (``1/2 $\chi^2$''). For $N_\sigma$ large enough one has: $TS\simeq N_\sigma^2$.  The $TS$ of the source is the sum of the contribution of each photon within the ROI with a  weight depending on $r_i$. As long as the ROI radius is large enough to include
all photons from the source, the $TS$ computed over the ROI is independent of the ROI size. Indeed, the contributions from distant photons are negligible since there are suppressed by the Point Spread Function.

\subsection{Relative flux uncertainty}

\noindent The relative flux uncertainty $\sigma_{\ln\,F}$ is calculated in Step 1 using the formulae below so as to consistently estimate $\sigma_{\ln\,F}$ as obtained in Step 2. These formulae are derived under the assumption that $\Gamma$ is left free in the fit performed in Step 2 (otherwise, one would just have  $\sigma_{\ln\,F}=\frac{\Delta A}{A}$).
  
The elements of the Hessian matrix, the inverse of the covariance matrix, can be calculated as: 
\begin{eqnarray}
H_{x_1 x_2} \equiv   \frac{\partial^2 \ln \mathcal{L}}{\partial x_1 \partial x_2}=   \frac{\partial^2 \Delta \ln \mathcal{L}}{\partial x_1 \partial x_2}
\label{eq:hess}
\end{eqnarray} 
where the parameters $x_{1,2}$ are either $A$ or $\Gamma$.
   
\noindent The relative statistical uncertainty on the differential flux is:
\begin{eqnarray}
\frac{\Delta A}{A}=\frac{|H_{A A}|^{-1/2}}{A}=  \left[\sum\limits_i \frac{g(r_i,E_i)^2}{(1+g(r_i,E_i))^2}\right]^{-1/2}
\label{eq:dAi}
\end{eqnarray} 
The cross-term element $H_{A \Gamma}$ is:
\begin{eqnarray}
H_{A \Gamma}=-\frac{1}{A}\left[\sum\limits_i \frac{g(r_i,E_i)\ln(E_i/E_0)}{(1+g(r_i,E_i))^2}-\iint Exp(E,T_0,T_1) S_S(E)PSF(r,E) \ln(E/E_0) d\Omega dE\right]
\end{eqnarray}
On average (over many observations of the same source), $H_{A \Gamma}$ can be written as 
\begin{eqnarray}
H_{A \Gamma}=\frac{1}{A} \int Exp(E,T_0,T_1) S_B(E) \ln(E/E_0) \int \frac{g(r,E)^2}{1+g(r,E)} d\Omega 
\end{eqnarray} 
assuming that the model is correct, i.e., the actual photon energy distributions are given by $S_S(E)$ and $S_B(E)$ for the source and background respectively.
Defining the pivot energy $E_{pivot}$ as the value of the reference energy  $E_0$ (Eq. \ref{eq:ss}) for which $H_{A \Gamma}=0$, one obtains:
\begin{eqnarray}
\ln(E_{pivot})= \frac{\int W(E) \ln E dE}{\int W(E) dE}
\label{eq:pivot}
\end{eqnarray} 
where
\begin{eqnarray}
W(E)= Exp(E,T_0,T_1) S_B(E) \int \frac{g(r,E)^2}{1+g(r,E)} d\Omega 
\label{eq:WE}
\end{eqnarray} 
In the following, the reference energy $E_0$  is set to $E_{pivot}$, as $\Delta A/A$ is minimum in that case. The Hessian matrix is then diagonal and thus so is the covariance matrix. 

\noindent $F$ is the integral flux between $E_{min}$ and $+\infty$. It can be written as:
\begin{eqnarray}
F=\int^{+\infty}_{E_{min}} S_S(E) dE=\frac{AE_{pivot}}{\Gamma-1}\left(\frac{E_{min}}{E_{pivot}}\right)^{1-\Gamma}
\label{eq:F}
\end{eqnarray}
The relative uncertainty of $F$, $\sigma_{\ln\,F}$, is obtained by propagating the errors of $A$ and $\Gamma$, making use of the fact that the covariance matrix is diagonal for this choice of $E_0$ : 
\begin{eqnarray}
\sigma_{\ln\,F}=\left[\left(\frac{\Delta A}{A}\right)^2+\sigma_\Gamma^2 \left(\ln\left(\frac{E_{min}}{E_{pivot}}\right)+\frac{1}{\Gamma-1}\right)^2\right]^{1/2}   
\label{eq:sigf}
\end{eqnarray}
where $\sigma_\Gamma$ is the uncertainty of $\Gamma$, $\sigma_\Gamma = |H_{\gamma \gamma}|^{-1/2}$.

\subsection{Optimum energy}

Using the optimum energy $E_1$, corresponding to the energy for which the second term in Eq. \ref{eq:sigf} cancels, for $E_{min}$ will lead to the minimum $\sigma_{\ln\,F}$ relative to other choices of $E_{min}$:
\begin{eqnarray}
\ln E_1=\ln E_{pivot} -\frac{1}{\Gamma-1}
\label{eq:decor2}
\end{eqnarray}
In that case $\sigma_{\ln\,F}$ is equal to $\Delta A/A$ and is obtained from Eq. \ref{eq:dAi}.

\end{appendix}

\end{document}